\pdfoutput=1
\documentclass[12pt]{article}
\usepackage[margin=0.8in]{geometry}
\usepackage{natbib}

\usepackage{comment}
\usepackage[percent]{overpic}
\usepackage[flushleft]{threeparttable}
\usepackage[clockwise, figuresright]{rotating}
\usepackage{pbox}
\usepackage{adjustbox}
\usepackage{multirow}
\usepackage{longtable}
\usepackage{enumitem}
\usepackage{rotating} 
\usepackage{array}
\newcolumntype{P}[1]{>{\centering\arraybackslash}p{#1}}
\usepackage{setspace}
\usepackage{authblk}
\usepackage[colorlinks=true,linkcolor=blue,allcolors=blue]{hyperref}   %% Comment for only arxiv submission
\usepackage{url}  %% Comment for only arxiv submission

\def\soho{{\sl SOHO}}
\def\stereo{{\sl STEREO}}
\def\sdo{{\sl SDO}}
\def\apj{Astrophys~J~}
\def\apjl{Astrophys~J~Lett~}
\def\grl{Geophys~Res~Lett~}

\def\apjs{\apj Suppl~}
\def\grl{Geophys~Res~Lett~}

\def\solphys{Sol~Phys~}

\def\jgr{J~Geophys~Res~}
\def\aap{Astro~and~Astrophys~}

\begin{document}

\title{Improving the Arrival Time Estimates of Coronal Mass Ejections by Using Magnetohydrodynamic Ensemble Modeling, Heliospheric Imager data, and Machine Learning}

\author[1]{Talwinder Singh}
\affil[1]{Center for Space Plasma and Aeronomic Research, The University of Alabama in Huntsville, AL 35805, USA}

\author[2]{Bernard Benson}
\affil[2]{McLeod Software Corporation, Birmingham, AL 35242, USA}

\author[3]{Syed A. Z. Raza}
\affil[3]{Department of Space Science, The University of Alabama in Huntsville, AL 35805, USA}

\author[1]{Tae K. Kim}

\author[1,3]{Nikolai V. Pogorelov}

\author[4]{William P. Smith}
\affil[4]{Department of Physics, Brown University, RI 02912, USA}

\author[5]{Charles N. Arge}
\affil[5]{Solar Physics Lab, NASA/GSFC, Greenbelt, MD 20771, USA}

\setcounter{Maxaffil}{0}
\renewcommand\Affilfont{\itshape\small}
\date{} 

%\begin{titlingpage}
\maketitle

\begin{abstract}
The arrival time prediction of Coronal mass ejections (CMEs) is an area of active research. Many methods with varying levels of complexity have been developed to predict CME arrival. However, the mean absolute error (MAE) of predictions remains above 12 hours, even with the increasing complexity of methods. In this work we develop a new method for CME arrival time prediction that uses magnetohydrodynamic simulations involving data-constrained flux-rope-based CMEs, which are introduced in a data-driven solar wind background. We found that, for 6 CMEs studied in this work, the MAE in arrival time was $\sim$8 hours. We further improved our arrival time predictions by using ensemble modeling and comparing the ensemble solutions with STEREO-A\&B heliospheric imager data. This was done by using our simulations to create synthetic J-maps. A machine learning (ML) method called the lasso regression was used for this comparison. Using this approach, we could reduce the MAE to $\sim$4 hours. Another ML method based on the neural networks (NNs) made it possible to reduce the MAE to $\sim$5 hours for the cases when HI data from both STEREO-A \& B were available. NNs are capable of providing similar MAE when only the STEREO-A data is used. Our methods also resulted in very encouraging values of standard deviation (precision) of arrival time. The methods discussed in this paper demonstrate significant improvements in the CME arrival time predictions. Our work highlights the importance of using ML techniques in combination with data-constrained magnetohydrodynamic modeling to improve space weather predictions.
\end{abstract}

\section{Introduction}\label{sec:Introduction}
Coronal Mass Ejections (CMEs) are coherent structures of plasma and magnetic field that originate from the Sun and propagate in to the interplanetary space through the ambient solar wind (SW). CMEs are major sources of geomagnetic storms at Earth, which have several undesirable consequences for many space- and ground-based technologies. Throughout the years, numerous observatories and instruments, both on the ground and in space, have assisted us in comprehending the connection between the Sun and Earth. Many CME propagation models have been developed over the years using the data from these observatories to provide CME arrival time forecasts. They include the empirical models \citep[e.g.][]{Vandas1996, Brueckner1998, Gopalswamy2001, Gopalswamy2005, Wang2002, Manoharan2004}, drag based models \citep{Vrsnak2001,Vrsnak2002}, and also such physics-based models as Shock Time of Arrival (STOA) \citep{Dryer1984}, STOA-2 \citep{Moon2002}, and HAFv.2 \citep{Fry2001}. Newer-generation prediction models are based on magnetohydrodynamic (MHD) simulations which use over-pressurized blobs to initiate the CME propagation  \citep[e.g., Wang--Sheeley--Arge (WSA)-ENLIL-Cone model,][]{Emmons13}). All these models result in arrival time mean absolute errors (MAEs) of more than 10 hours. \citet{Riley2018} compared several CME arrival time prediction models and found that even the best performers had  MAEs of 13 hours and standard deviation of 15 hours. It was also shown that the arrival time accuracy of the models is not improving with the model complexity \citep{Riley2018}. This has been demonstrated by other studies \citep[e.g.,][]{Gopalswamy2013}, which showed that the performance of the ENLIL-cone model is similar to that of the much simpler empirical shock arrival model \citep{Gopalswamy2005}. In addition, all such models cannot predict the north-south magnetic field component ($B_z$) in the CME at Earth because they ignore the development of the associated magnetic flux ropes.

CMEs disturb the ambient SW during their travel through the inner heliosphere (IH). Reciprocally, the SW can impact the kinetic and magnetic properties of CMEs. Understanding this CME-SW interaction is important for accurate forecasting of the arrival time of CMEs. With the availability of Solar Terrestrial Relations Observatory (STEREO) Heliospheric Imager (HI) data \citep{Eyles09}, it became possible to track the kinematics of CMEs up to large distances from the Sun \citep{Liu10, Lugaz10} and study the CME-SW interaction with more detail. HI data have been used to develop several arrival time forecasting methods. These methods track CME fronts in the ecliptic plane using the HI data. The {pixels corresponding to the ecliptic plane} are identified in the running-difference HI images. {These pixels} are arranged in a vertical formation over a specified period of time, such that time forms the horizontal axis, to create time-elongation maps, also called J-maps. CME fronts reveal themselves as diagonal bright streaks in J-maps. To calculate the distance the CME has traveled along the Sun-Earth line using these J-maps, several models have been developed such as the fixed $\phi$ \citep{Kahler07}, harmonic mean fitting \citep{Lugaz09}, triangulation with harmonic mean \citep[HM,][]{Lugaz10}, and self-similar expansion \citep[SSE,][]{Davies12} models. All of them use very simple assumptions about the shape of a CME. Both HM and SSE models assume CMEs to have a spherical shape, with the sphere either connected to the solar center (in the HM model) or expanding and propagating in the IH (in the SSE model). Forecasting methods that use these models usually extract CME time-elongation data from the J-maps to where the bright fronts are traceable and then use drag-based models to forecast arrival time at Earth. These methods, however, have not been able to surpass the accuracy of such models as the WSA-ENLIL-Cone model, which use only the coronagraph data to forecast CMEs. Simplistic assumptions about initial CME shapes in the IH and ignorance of the CME--SW interactions, which can distort their shapes during propagation can be a reason for this \citep{Liu06, riley97, Manchester04}. 

The EIEvoHI model, first described by \citet{Rollett16} assumes a more flexible CME shape while converting the CME-front elongation angle to the distance a CME has traveled along the Sun-Earth line. That model assumes CMEs to have elliptic shape with a flexible ellipse aspect ratio. It also tracks CME fronts in HI  data to a particular distance and uses a drag-based model to find the CME arrival time at Earth. Ensemble modeling with EIEvoHI has been shown to be a promising method of CME arrival time predictions \citep{Amerstorfer17}. A recent study by \citet{Hinterreiter21} reported MAE of 7.5 hours and a standard deviation of 9.5 hours by hind-casting 12 CMEs using EIEvoHI. EIEvoHI is only suitable for forecasting the arrival time and speed of CMEs at Earth. It does not consider the associated flux ropes, so it is unable to estimate the magnetic field carried by a CME towards Earth.

MHD modeling can reproduce realistic SW-CME interaction and eliminate the necessity of making assumptions about the CME shape in the IH. It provides us with a promising tool for forecasting the properties of the SW and CME plasma, and magnetic field at Earth. This requires that the solar wind is driven by data and CMEs  simulated on the basis of flux rope models \citep{Jin17b, Scolini19, Singh20b, Singh22}. Flux rope models are usually constrained by CME observations in coronagraphs. \citet{Scolini19} and \citet{Singh19} constrained the poloidal flux in their models using the reconnected flux in the source active regions. \citet{Singh22} introduced a constant turn flux rope model and used it to simulate the 12 July 2012 CME. This model can be constrained by the CME direction, tilt, speed, half-angle, aspect ratio, magnetic flux, and helicity sign. \citet{Singh22} reproduced the plasma and magnetic field properties of this CME at Earth with good accuracy. They also described a method to create ensemble members for this flux-rope-based CME and used them to study the role of observational uncertainties in the CME properties obtained at Earth with numerical simulations. HI observations can be used to compare the evolution of different ensemble members with the actual CME behavior. This type of comparison was done earlier by \citet{Barnard20} to improve the arrival time accuracy of their CME model. 

In this work, we use the method described by \citet{Singh22} to perform an MHD, ensemble modeling of six CMEs. We propose a novel approach to comparing our ensemble members with observations in the IH using ML methods. We show that incorporating HI data in MHD forecasting methods can substantially improve the arrival time predictions. We describe the data used in this work in Section~\ref{sec:Data}. In Section~\ref{sec:Models}, we describe the SW model and CME models used in this study. We show our results in Section~\ref{sec:Results}, which is followed by our conclusions in Section~\ref{sec:Conclusions}.

\section{Data Used}\label{sec:Data}
In this work we analyze six CMEs listed in Table~\ref{table:1}. For each CME, we use the level~0.5 FITS data from the Sun-Earth Connection Coronal and Heliospheric Investigation (SECCHI)/Cor2 ~\citep{Howard08} coronagraphs in Solar Terrestrial Relations Observatory (\stereo) A \& B~\citep{Kaiser08}. These files are processed to level~1 using the \textit{secchi\_prep} program in IDL's SolarSoft library which converts the units from data numbers (DN) into Mean Solar Brightness (MSB). We also use level~1 FITS data from Large Angle Spectroscopic Coronagraph (LASCO)/C2/C3~\citep{Brueckner95} coronagraph in the Solar and Heliospheric Observatory (\soho) spacecraft. The coronagraph data are used to estimate the direction, tilt, half-angle, aspect ratio, and speed of the CMEs. We use long-term background-subtracted data from HI 1 \& 2 \citep{Eyles09} onboard \stereo\  A \& B to track the CMEs in the IH. We also use level 2 HI data which are already corrected for cosmic rays, shutterless readout, saturation effects, flat fields, and instrument offsets from spacecraft pointing. We also use the solar extreme ultraviolet (EUV) and line of sight (LOS) magnetic field observations from the Solar Dynamics Observatory (\sdo) Atmospheric Imaging Assembly (AIA) \citep{Pesnell12} and Helioseismic and Magnetic Imager (HMI) \citep{Schou12, Hoeksema14} instruments to study, respectively, the source regions of the six CMEs. We use \textit{aia\_prep} program in SolarSoft to read in and calibrate AIA level~1 data to level~1.5. Level~1 data includes bad-pixel removal, de-spiking, and flat-fielding. Level~1.5 data additionally has roll correction that makes the solar north in the images vertical, re-scales the images to $0''.6$  pixels, and performs translation that puts the solar disk center at the image center. The \textit{hmi\_prep} program in SolarSoft is utilized to read HMI data. This program processes the magnetograms to guarantee the proper roll angle and adjusts their placement so that the center of the solar disk aligns with the centers of the magnetograms. AIA and HMI data of the source active regions are used to estimate the magnetic flux and helicity sign of the six CMEs. The methods used to estimate these values are described in Section~\ref{sec:Results}. One hour averaged data, provided by NASA/GSFC's OMNI data through OMNIWeb \citep{King05}, are used to compare the in situ measurements at Earth with our simulations.  All CMEs we study in this work are Earth-directed and show magnetic cloud signatures at Earth. The CME speeds near the Sun are in a wide range between 500 and 1500 km/s.

\section{Simulation Models}\label{sec:Models}
We perform MHD simulations using the Multi-Scale Fluid-Kinetic Simulation Suite (MS-FLUKSS) in this study. MS-FLUKSS is a collection of modules capable of performing adaptive mesh refinement (AMR) simulations of the SW in the presence of neutral atoms, non-thermal ions, turbulence, etc. \citep{Pogorelov14, Fraternale21}. This code is highly parallel, allowing us to perform simulations much faster than in real time. We describe the SW and CME models used in this study in the following subsections.

\subsection{Solar wind model}
The IH model implemented in MS-FLUKSS solves the ideal MHD equations with the finite-volume, total variation diminishing (TVD) approximations of conservation laws on non-uniform spherical grids. For this study, we chose the grid with 150, 256, and 128 cells in the $r$, $\phi$, and $\theta$ directions, respectively. We set the inner boundary of our domain at 0.1 AU, where the SW has already attained super-fast magnetosonic speed. The outer boundary is set at 1.5 AU. The inner boundary conditions are provided as a time-series of WSA solutions \citep[e.g.,][]{Kim19}. The WSA results at this height are obtained using the Air Force Data Assimilative Photospheric Flux Transport (ADAPT) synchronic maps based on the NSO/GONG magnetograms \citep{Arge10, Arge11, Arge13, Hickmann15} as input at the solar surface. The WSA model employs the potential field source surface (PFSS) model to extend the photospheric magnetic field to a spherical source surface located at 2.5 $R_\odot$ and utilizes the Schatten current sheet model \citep[PFCS,][]{Schatten71} to take it even further to the outer boundary of the WSA at 0.1 AU. The SW speed at the WSA outer boundary is calculated as a function of the flux expansion factor and distance to the nearest coronal hole boundary \citep{Arge03, Arge05, McGrogor2011}. SW density and temperature at 0.1 AU are estimated by using the empirical correlations between the SW speed, density, and temperature based on the OMNI data \citep{Elliott16}. The process of using WSA boundary conditions in the IH model is the same as described in \citet{Singh20b} and \citet{ Singh22}. ADAPT-WSA provides 12 realizations that can be used as the inner boundary in our IH model. To perform our CME simulations, we use the best ADAPT-WSA realization based on their comparison with near-Earth SW data. 

\begin{figure}
\center
\includegraphics[scale=0.1,angle=0,width=8cm,keepaspectratio]{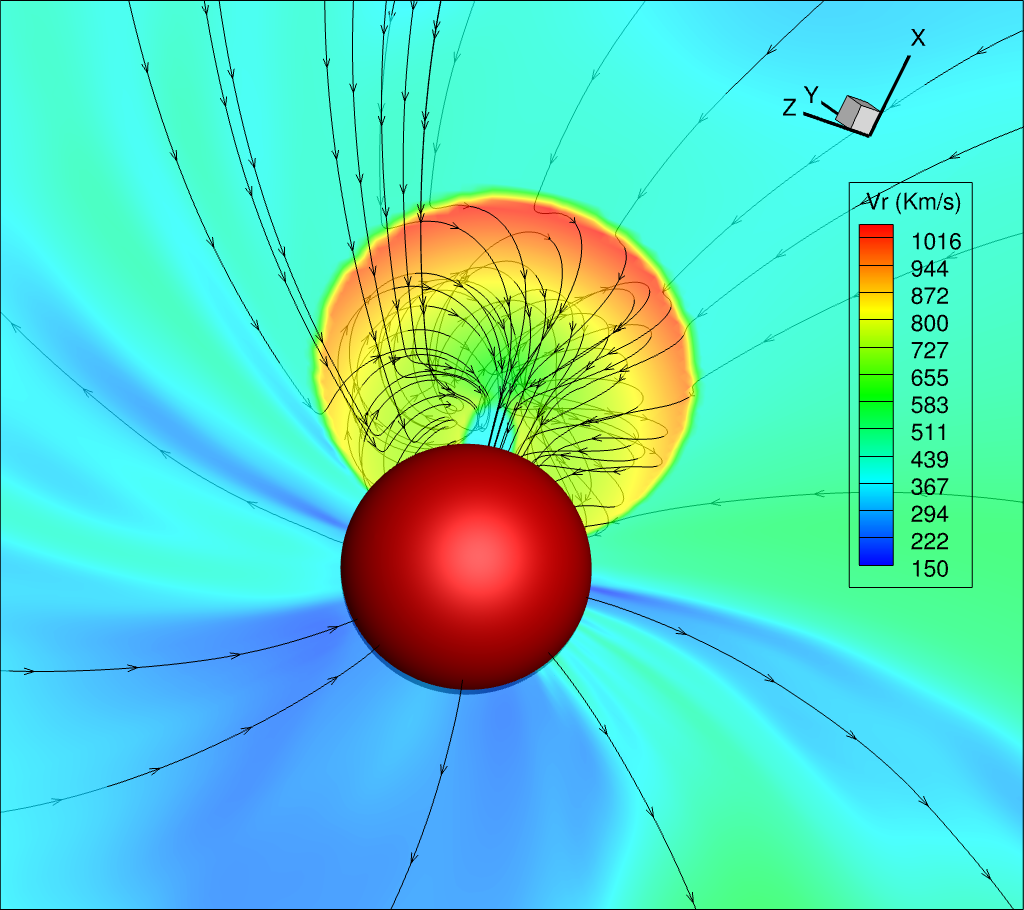}
\caption{The initial structure of the constant turn flux rope used to simulate the 12 July 2012 CME (case 5). The speed at the apex is 1097 km/s, and the poloidal and toroidal magnetic fluxes are set to $1.4\times10^{22}$ Mx and  $7.6\times10^{21}$ Mx respectively. The flux rope is inserted into the IH at the physical time of 13 July 2012 at 03:33. The apex of the flux rope is at $70R_\odot$, while its latitude, longitude, and tilt are $\mathbf{-10.1^\circ}$, $\mathbf{0.7^\circ}$, and $\mathbf{52.0^\circ}$ respectively. The semi-translucent slice by the plane containing the flux rope axis shows the radial velocity. Magnetic field lines are shown with black lines. The red sphere indicates the inner boundary of the IH model at 0.1 AU.}
\label{Initial_Vr}
\end{figure}
\subsection{CME model}
In this work, we use a constant turn flux rope model described in \cite{Singh22} to simulate CMEs. This model assumes the initial CME structure to be croissant-like with a circular cross-section and two legs rooted at the center of the Sun. This shape is based on FRiED model geometry described by \citet{Isavnin16}. The magnetic field of this flux rope model is based on the analytic solutions provided by \citet{Vandas17}. The constant turn flux rope can be initialized in any direction defined by the apex latitude and longitude. The flux rope can have any tilt w.r.t. the solar equatorial plane. Moreover, this flux rope can have desired half-angle and aspect ratio. The magnetic field in the flux rope can be initialized to have desired poloidal and toroidal fluxes, and helicity sign. The helicity sign can take the +1 or -1 values and describes the direction of the twist of  magnetic field lines in the flux rope, with the +1 (-1) sign representing the direction described with the right- (left-) hand rule. The velocity of plasma inside the flux rope is initialized to have desired value at the apex and a self-similar expanding profile \citep{Singh22}. We can introduce a uniform-density plasma inside this flux rope. \citet{Singh22} showed that the constant turn flux rope model is suitable for simulating CMEs in the IH. It reproduced the CME magnetic field, density, and velocity reasonably well at Earth for the 12 July 2012 CME, when this model was constrained with the observed direction, tilt, half-width, aspect ratio, speed, mass, helicity sign, and magnetic flux. These properties of the CMEs can be obtained using various observational data as discussed in \citet{Singh20a}. This CME is also included in the current work as case 5.

Flux ropes are inserted into the ambient SW in such a way that the former are initially superimposed with the SW background \citep{Singh20b}. When a flux rope is superimposed with the background, we replace the SW magnetic field and velocity with those in the flux rope. During this superimposition, The density and internal energy density of the flux rope are added to the values in the ambient SW density and internal energy. Figure~\ref{Initial_Vr} shows the initialized flux rope for the simulation of 12 July 2012 CME. The properties of this flux rope are summarised in Table~\ref{table:1}. We will describe the method of constraining this model with observations in Section~\ref{subsec:Constrain}.

\section{Results}\label{sec:Results}

\subsection{Constraining initial CME parameters with data}\label{subsec:Constrain}
The CME properties that we need to constrain our flux rope are its direction, tilt, speed, half-angle, aspect ratio, mass, poloidal and toroidal magnetic fluxes, and the magnetic helicity sign. These properties for our six cases are summarized in Table~\ref{table:1}. These quantities are estimated using different observations as discussed below. 

We can estimate the CME direction, tilt, speed, half-angle, aspect ratio, and speed using multi-viewpoint observations of STEREO-A \& B {COR2} and SOHO C2/C3 coronagraphs. A suitable model for utilizing these multi-viewpoint observations to find 3D CME evolution is the Graduated Cylindrical Shell (GCS) model \citep{Thernisien11}. This model fits a croissant shape with a curved front and conical legs over a CME as seen in the coronagraph field of view (FOV). Table~\ref{table:1} shows the fitted latitude, longitude, tilt angle from the solar equatorial plane, half-angle, and aspect ratio for the six CMEs we have considered in this study. Each time series of GCS fitted heights gives us the CME speed via linear regression \citep[see e.g.][]{Hess14}. This value has also been reported in Table ~\ref{table:1}. It should be noted that the GCS model can work only when a CME is in the coronagraph FOV. However, we insert our flux rope into IH when its apex is already at $70R_\odot$, by which time the CMEs have already left the STEREO and SOHO coronagraph FOVs. We assume a self-similar expansion of CMEs between the time we are last able to fit the GCS model and when the CME apex should reach $70R_\odot$ according to the drag-based model (DBM) \citep{Vrsnak07}. We used the drag parameter of $10^{-8}$  \textup{${km^{-1}}$} and the asymptotic SW speed equal to 450 km/s for all our cases. \citet{Vrsnak14} found that this combination of SW speed and drag parameter ensures roughly the same arrival time accuracy for the DBM and WSA-ENLIL-Cone models. DBM also estimates the CME speed at $70R_\odot$ (see Table~\ref{table:1}). We superimpose each flux rope with the SW background at this DBM--estimated time and assume the DBM--estimated speed at the apex.

The total mass of a CME can be found using the CME brightness in the white light coronagraph images. The brightness of a CME in coronagraphs is due to the Thomson scattering of photospheric light by the CME plasma electrons \citep{Billings66}. We can calculate the true mass of a CME by integrating the brightness over the CME area in coronagraph images and removing the projection effects, which is done with the method based on the multiple coronagraph viewpoints \citep{Colaninno09}. These true masses of the six CMEs under consideration are given in Table~\ref{table:1}. In this study, we uniformly distribute 0.1\% of this mass inside the flux rope when initializing it. The motivation behind using this factor is based on the investigation of \citet{Singh22} who showed that this results in a reasonable reproduction of the observed plasma density at Earth for 12 July 2012 CME. The physical interpretation of this factor being so small is that only the flux-rope part of a CME is inserted into the ambient SW in our simulation. The coronagraph observations show that flux rope resides in the cavity region of a CME and has density much lower than those in the sheath and core, which contribute the majority of mass to CMEs \citep{Riley08}. 

We estimated the poloidal magnetic flux of our CMEs by using the reconnected flux in post-eruption arcades (PEAs) in the source active regions \citep{Gopalswamy18}. This technique necessitates EUV observations when the PEA structure has reached its full development, usually during the decline stage of the flare.. \citet{Gopalswamy17} showed that the poloidal flux of CME flux ropes is equal to one half of the unsigned flux in the area covered by the PEAs at this time. In this paper, we estimate the toroidal fluxes of our CMEs by using an analytic relation between the poloidal and toroidal magnetic fluxes in CME flux ropes given by \citet{Qiu07}. The poloidal and toroidal fluxes found by using these methods for our six CMEs are given in Table~\ref{table:1}. The toroidal flux can also be estimated by using the coronal dimming area in the EUV observations at the time of eruption. However, coronal dimming is not apparent for many CME eruptions, so the toroidal flux estimates based this method are possible only {for} a limited number of CMEs. Therefore, we opted to use the analytic relationship by \citet{Qiu07}. The constant turn flux rope model can be initialized with desired poloidal and toroidal fluxes. This is an advantage of this model compared to the spheromak model where the poloidal and toroidal fluxes cannot be set independently. 

The sign of the helicity in a CME flux rope can be established by examining the pre-eruptive magnetic field configuration of the active region, as reported by \citet{Bothmer98}. For instance, \citet{Luoni11} illustrate how the magnetic tongues present in active regions can be used to determine the helicity sign of the flux ropes above them. We use this method to find the helicity sign for our six cases (see Table~\ref{table:1}). The constant turn flux rope model can be initialized to have either right-handed or left-handed magnetic field turns based on the helicity sign found by using this method. The hemispheric helicity rule can also be used to determine the helicity sign of the ARs, which states that ARs in the solar northern (southern) hemisphere have negative (positive) helicity. However, \citet{Liu14} showed that this rule did not apply to about a quarter of the 151 ARs studied by them. We found that all six CMEs discussed here follow the hemispheric helicity rule.

\begin{table}
\centering
\caption{Properties of the six CMEs discussed in this study.\label{table:1}}
\begin{tabular}{|P{2.3cm}|P{2cm}|P{2cm}|P{2cm}|P{2cm}|P{2cm}|P{2cm}|}
\hline
CME \#  & 1 & 2 & 3 & 4 & 5 & 6 \\ \hline
Date  & 2010/08/01 08:00 & 2011/09/06 23:05 & 2011/09/13 22:10 & 2012/01/19 14:36 & 2012/07/12 16:48 & 2012/09/28 00:00 \\ \hline
Latitude & 17.9 & 26.8 & 21.2 & 43 & -10.1 & 11.7 \\ \hline
Longitude& -37.9 & 40 & 17.6 & -30.4 & 0.7 & 31.3 \\ \hline
Tilt & 75.5 & 16.2 & -11.2 & -57 & 52 & -78.3 \\ \hline
Half angle & 35 & 55 & 30 & 43 & 33 & 40 \\ \hline
Ratio & 0.5 & 0.5 & 0.5 & 0.5 & 0.5 & 0.5 \\ \hline
Linear speed  & 1441 & 746 & 547 & 1214 & 1267 & 976 \\ \hline
DBM speed & 1218 & 708 & 541 & 1061 & 1097 & 887 \\ \hline
Mass ($\times10^{15}$ g)& 3.8 & 10.3 & 4.5 & 18.1 & 16.5 & 9.6 \\ \hline
Poloidal Flux ($\times10^{21}$ Mx)& 1.7 & 2.9 & 3.6 & 3.7 & 14.0 & 0.9\\ \hline
Toroidal flux ($\times10^{21}$ Mx)& 0.6 & 1.2 & 1.5 & 1.6 & 7.6  & 0.3\\ \hline
Helicity & -1 & -1 & -1 & -1 & 1 & -1 \\ \hline
\end{tabular}
\end{table}

\subsection{Ensemble modeling}
The observations described in the previous subsection are not free from uncertainties. These uncertainties can be due to measurement errors, model assumptions, and subjective errors. \citet{Singh22} focused on the uncertainties in the GCS fitting parameters and found the subjective uncertainties in CME latitude, longitude, tilt, and speed by comparing the fitting results of 56 CMEs reported in multiple studies and catalogs. They found that the GCS estimates of the CME latitude, longitude, tilt, and speed have average uncertainties of $5.7^\circ$, $11.2^\circ$, $24.7^\circ$, and 11.4\% respectively. \citet{Thernisien09} have reported the average uncertainties in aspect ratio and half-angle GCS parameters as $+0.07/-0.04$ and $+13^\circ/-7^\circ$ respectively. Using the GCS uncertainties, \citet{Singh22} created 77 ensemble members of the 12 July 2012 CME and analyzed how the GCS uncertainties can create a spread in the simulated CME properties at Earth.

In this study, we similarly create 77 ensemble members for each of our six cases. The first member is called the seed ensemble member and the other 76 are created by introducing uncertainties to its GCS parameters in various combinations. The seed ensemble member with the GCS properties of latitude $\theta$, longitude $\phi$, tilt $\gamma$, speed $V_\mathrm{CME}$, half-angle $\alpha$, and aspect ratio $\kappa$ can be represented by a set $\{\theta, \phi, \gamma, V_\mathrm{CME}, \alpha, \kappa\}$. The other 76 ensemble members are created by introducing the average GCS uncertainties to these properties. These members are represented by the following sets:
\begin{enumerate}[noitemsep,nolistsep]
    \item$\{\theta\pm 5.7^\circ, \phi\pm 11.2^\circ, \gamma\pm 24.7^\circ, V_\mathrm{CME}\pm 11.4\%, \alpha_{-7^\circ}^{+13^\circ}, \kappa_{-0.04}^{+0.07} \}$, (64 members)
    \item$\{\theta\pm 5.7^\circ, \phi, \gamma, V_\mathrm{CME}, \alpha, \kappa\}$ (2 members)
    \item$\{\theta, \phi\pm 11.2^\circ, \gamma, V_\mathrm{CME}, \alpha, \kappa\}$ (2 members)
    \item$\{\theta, \phi, \gamma\pm 24.7^\circ, V_\mathrm{CME}, \alpha, \kappa\}$ (2 members)
    \item$\{\theta, \phi, \gamma, V_\mathrm{CME}\pm 11.4\%, \alpha, \kappa\}$ (2 members)
    \item$\{\theta, \phi, \gamma, V_\mathrm{CME}, \alpha_{-7^\circ}^{+13^\circ}, \kappa\}$ (2 members) 
    \item$\{\theta, \phi, \gamma, V_\mathrm{CME}, \alpha, \kappa_{-0.04}^{+0.07}\}$ (2 members)
\end{enumerate}

Ensembles for each of our six CMEs consist of 12 realizations provided by ADAPT-WSA. These are used as boundary conditions to obtain the SW properties at Earth. To perform a CME simulation, we choose a realization that provides the best performance at Earth for the ambient SW. Currently we use a manual approach to select the best-performing ADAPT-WSA ensemble member. For each of our six CMEs, we conduct 77 simulations by initializing the flux ropes using the CME properties listed in Table~\ref{table:1} for the simulation of the seed ensemble member and using the method described previously to create the remaining 76 ensemble members. In this work, we simulate ensemble members with apex speeds at $70R_\odot$ as low as 480 km/s and as high as 1357 km/s. The simulated CMEs had poloidal fluxes ranging between $0.9\times10^{21}$ Mx and $14.0\times10^{21}$ Mx. We simulate ensemble members that have CME half angles ranging between $23^\circ$ and $68^\circ$. This demonstrates the ability of our constant turn flux rope model to simulate CMEs with a wide range of speeds, magnetic fluxes, and widths.

\begin{figure}
\centering
\center
\begin{tabular}{c c} 
\begin{overpic}[scale=0.1,angle=0,width=7cm,keepaspectratio]{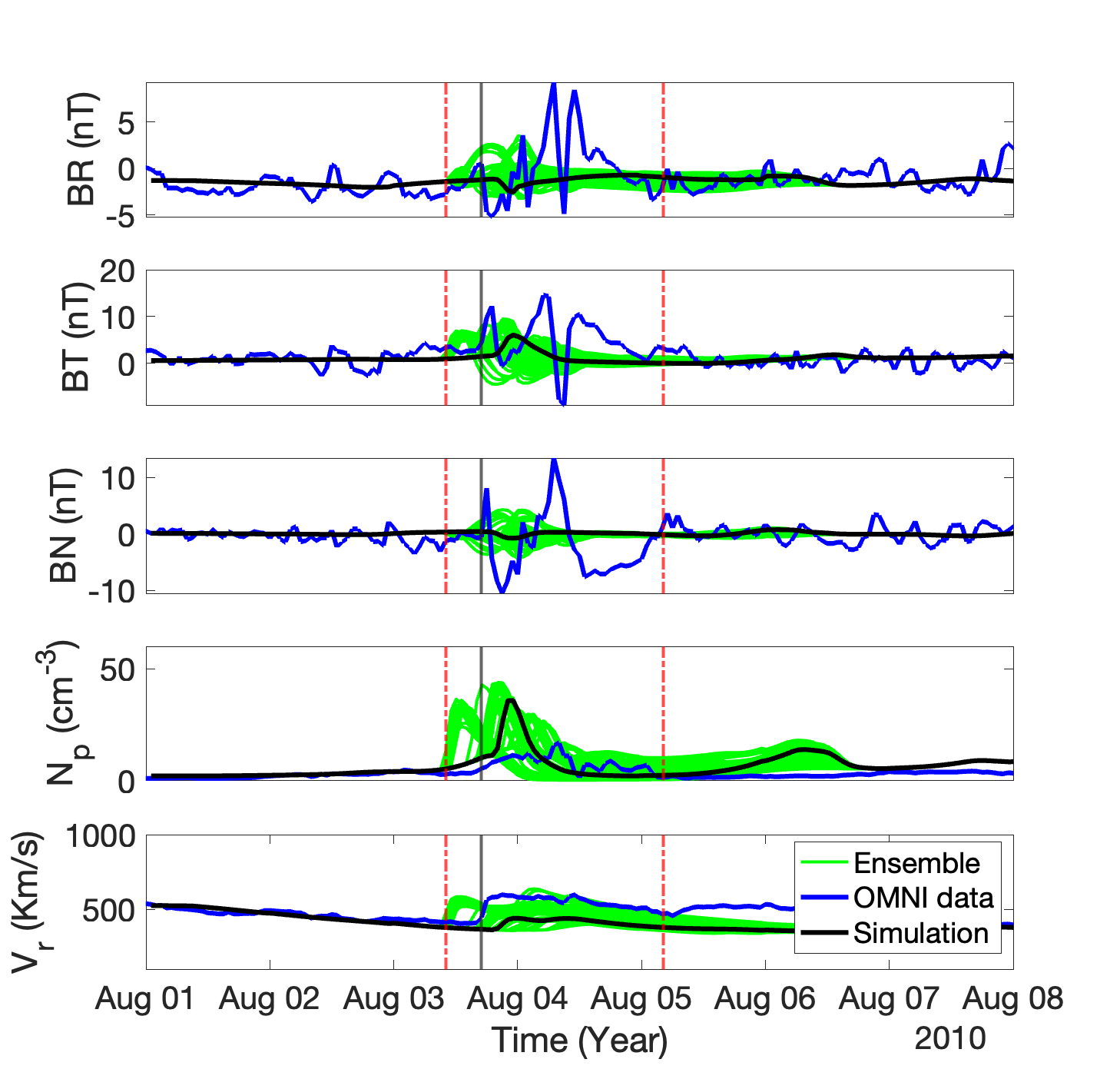}
\put(82.0,92){\color{black}{ \fontsize{8}{9}\selectfont 1}}
\end{overpic}
\begin{overpic}[scale=0.1,angle=0,width=7cm,keepaspectratio]{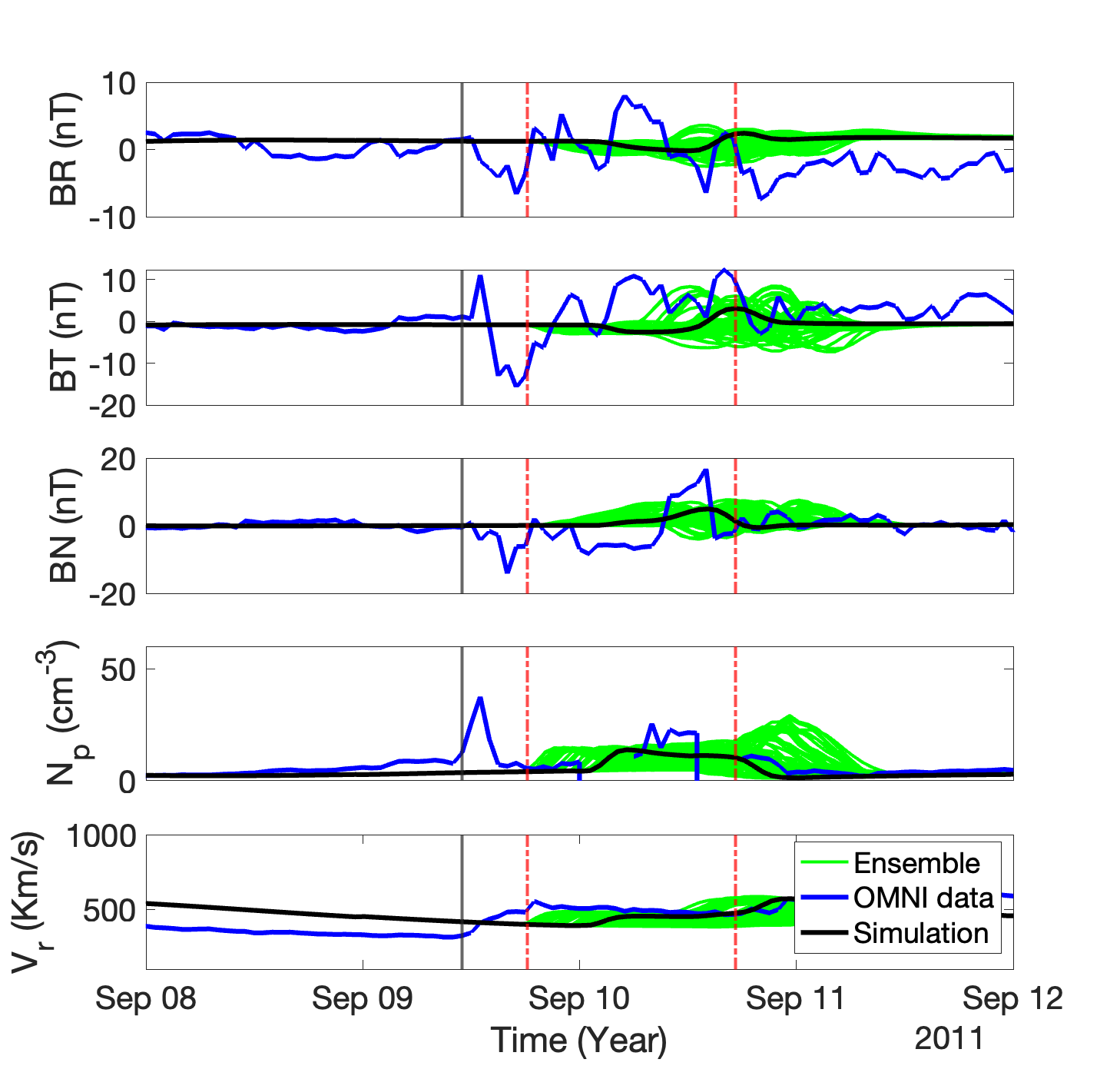}
\put(82.0,92){\color{black}{ \fontsize{8}{9}\selectfont 2}}
\end{overpic}\\
\begin{overpic}[scale=0.1,angle=0,width=7cm,keepaspectratio]{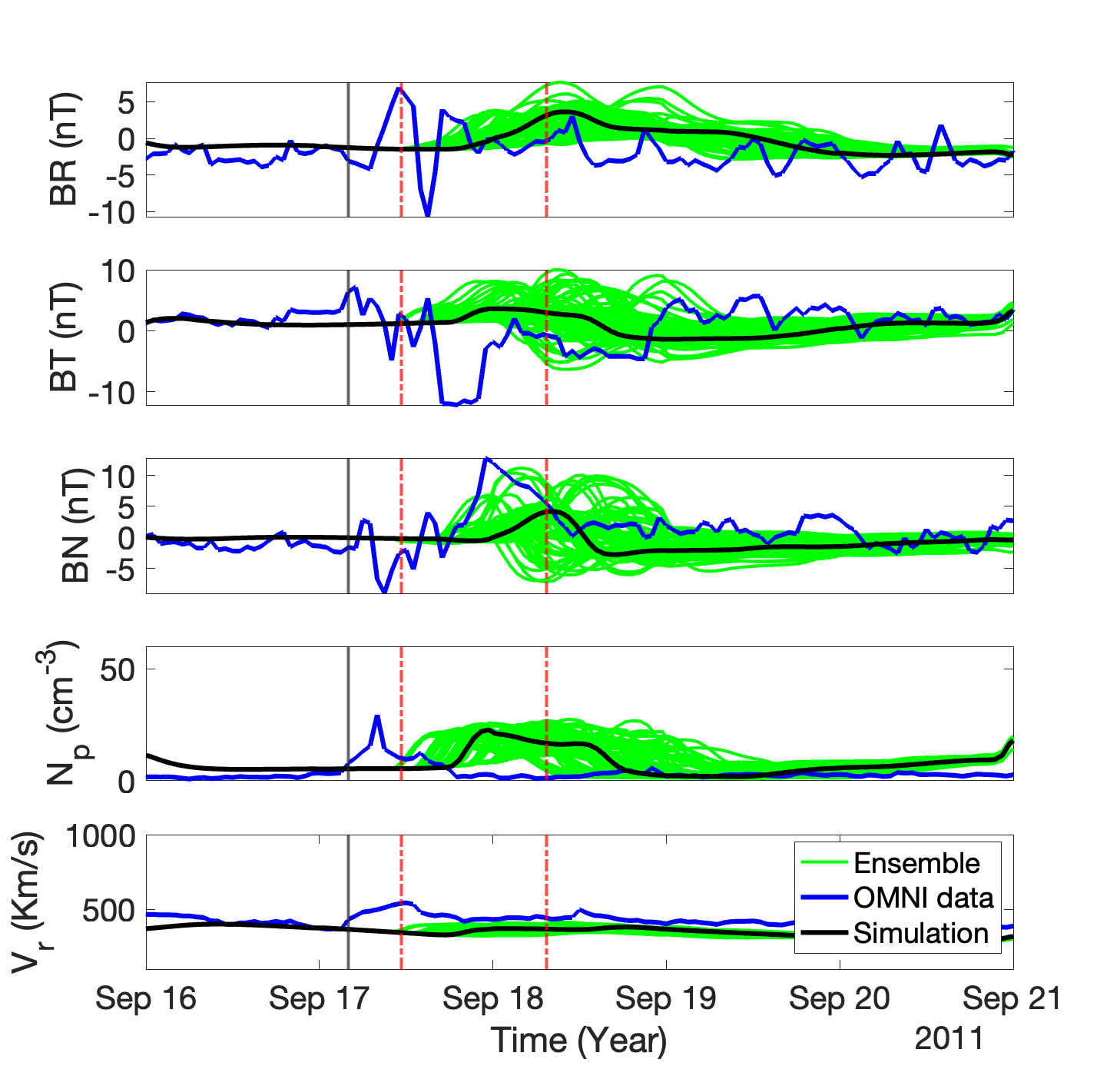}
\put(82.0,92){\color{black}{ \fontsize{8}{9}\selectfont 3}}
\end{overpic}
\begin{overpic}[scale=0.1,angle=0,width=7cm,keepaspectratio]{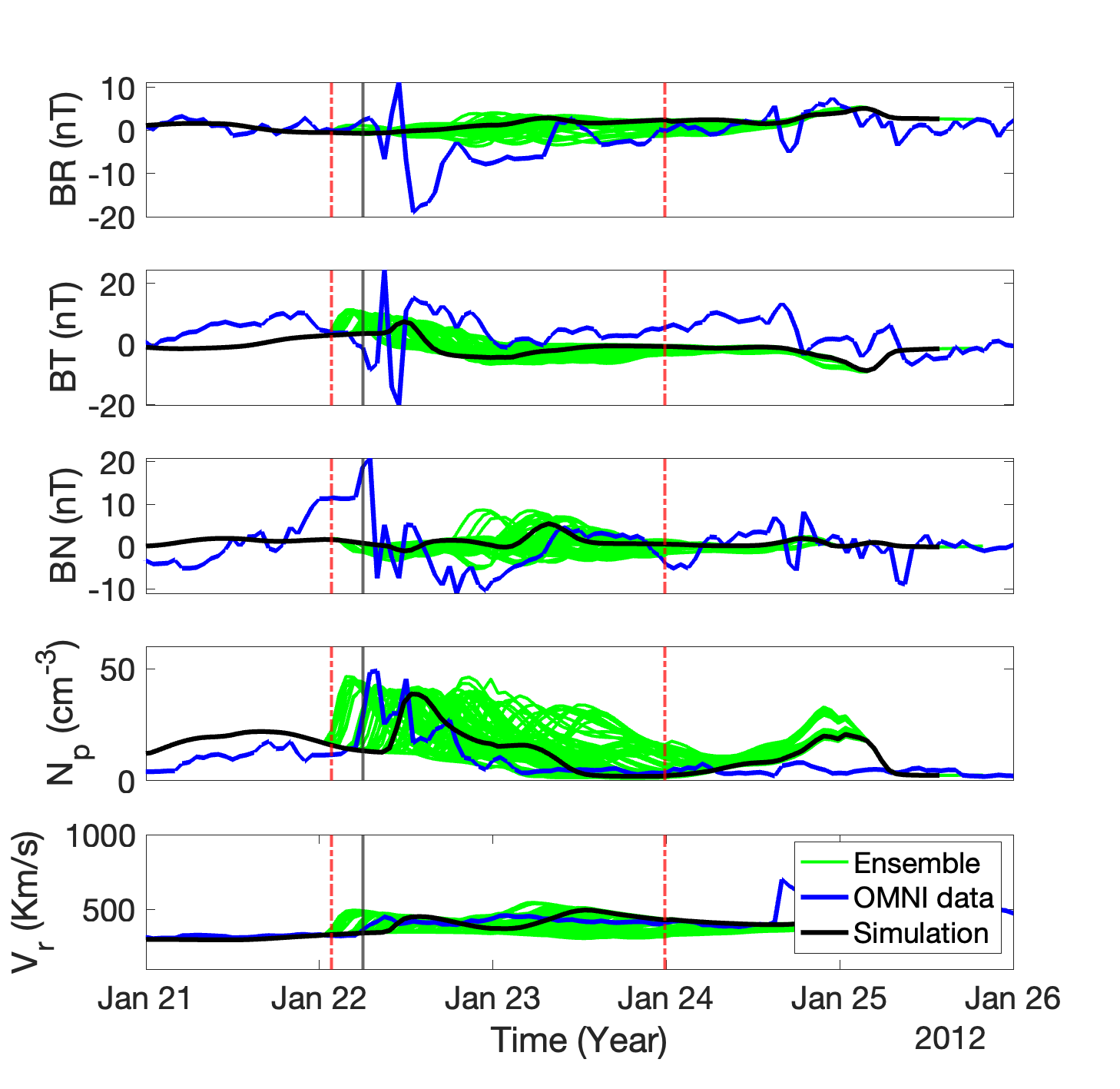}
\put(82.0,92){\color{black}{ \fontsize{8}{9}\selectfont 4}}
\end{overpic}\\
\begin{overpic}[scale=0.1,angle=0,width=7cm,keepaspectratio]{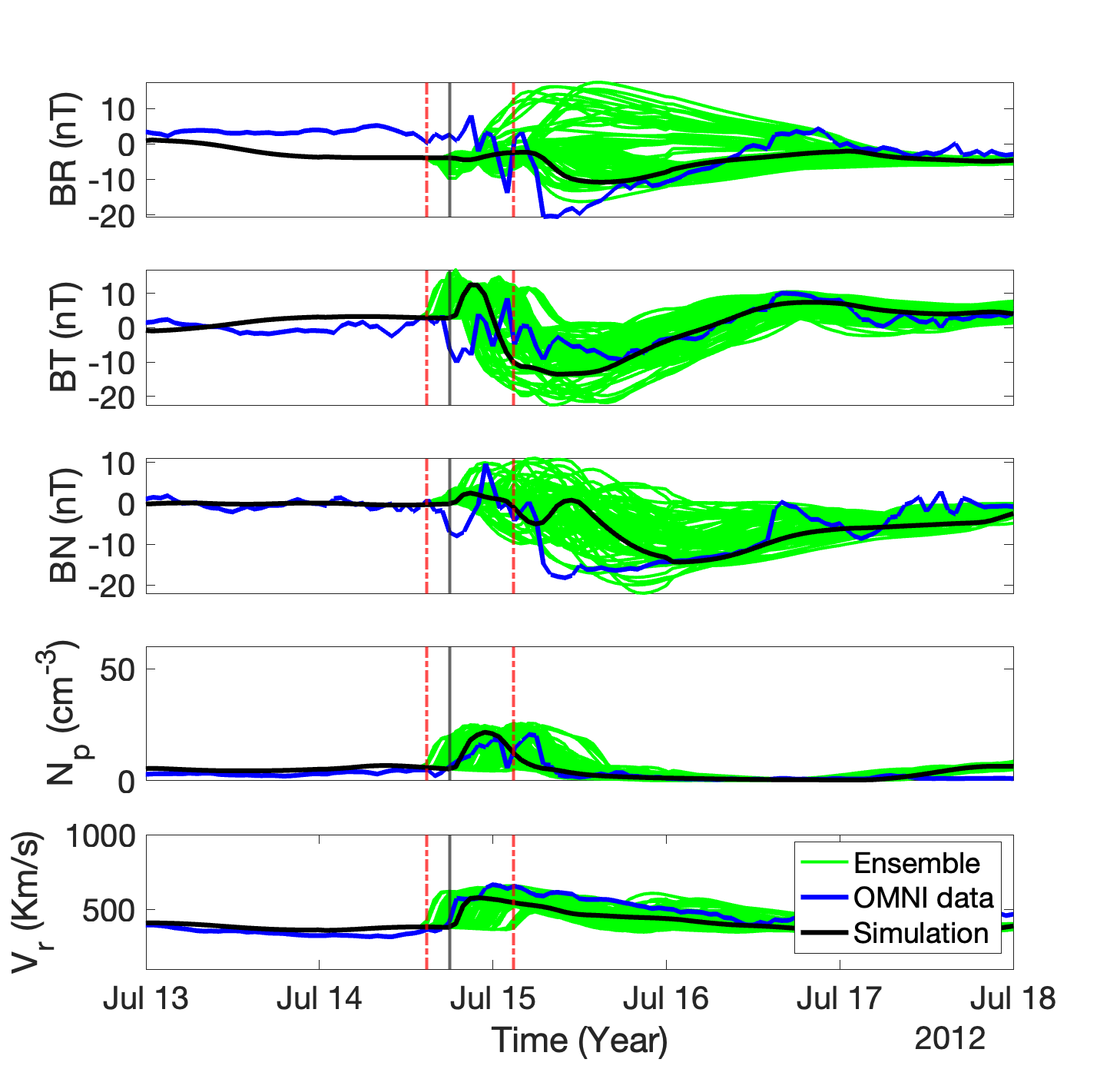}
\put(82.0,92){\color{black}{ \fontsize{8}{9}\selectfont 5}}
\end{overpic}
\begin{overpic}[scale=0.1,angle=0,width=7cm,keepaspectratio]{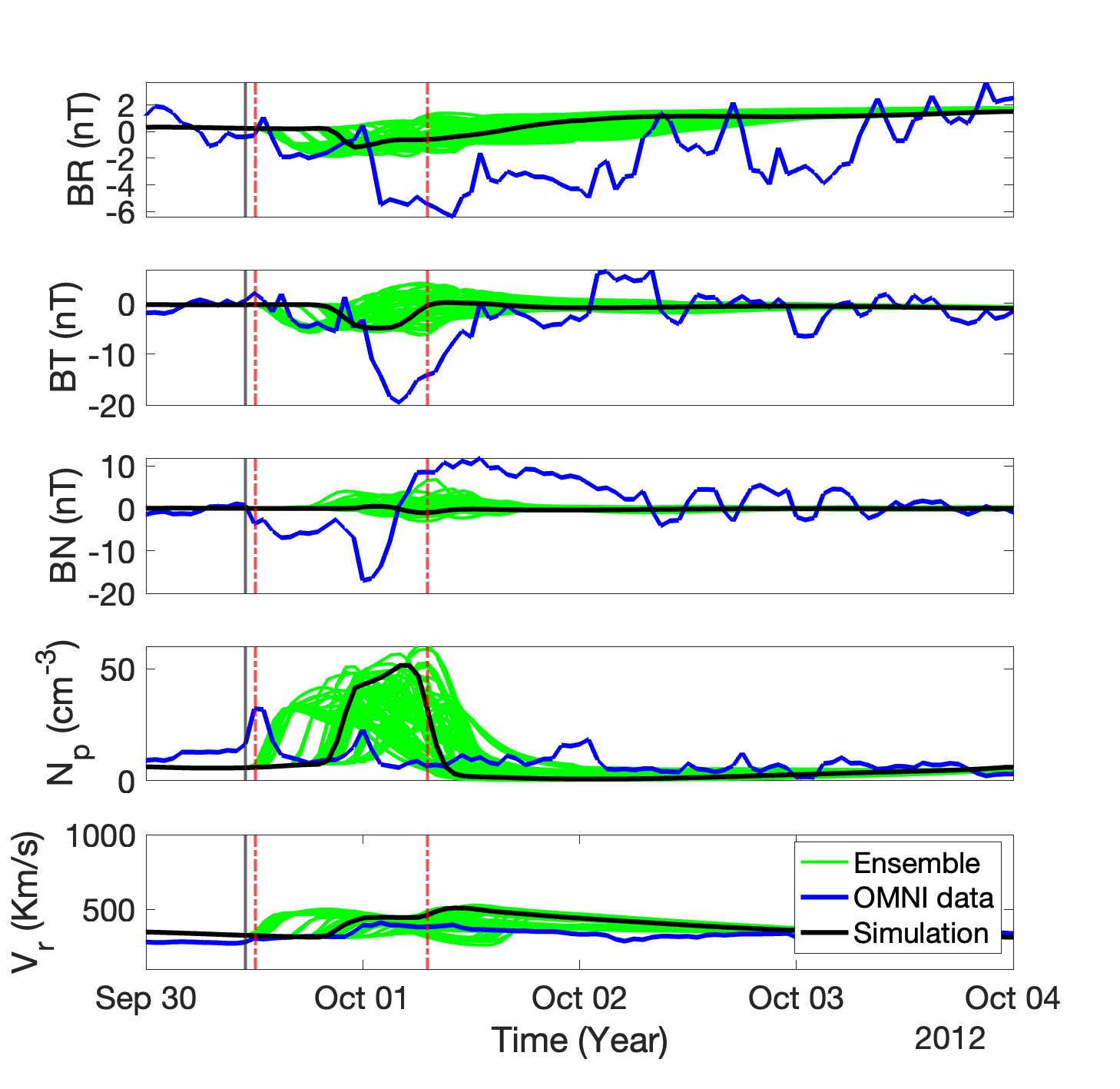}
\put(82.0,92){\color{black}{ \fontsize{8}{9}\selectfont 6}}
\end{overpic}
\end{tabular}
\caption{Plasma density, radial velocity components, and the heliospheric magnetic field in the 1 hr averaged OMNI data (blue lines) and our simulation results at Earth (black and green lines) for Cases 1 to 6. The case numbers are shown beside the panels. The black lines represent the results obtained when using the seed ensemble members. The green lines show the results for the 76 ensemble members, which were found by introducing uncertainties in the GCS parameters of the seed ensemble member. {The vertical gray line signifies the arrival of the observed CME, while the dotted red lines indicate the arrival times of the earliest and latest simulated CME ensemble members.}}
\label{1AU_data}
\end{figure}

Figure~\ref{1AU_data} shows the CME properties at Earth for all six cases. Each panel has been tagged with the CME number given in Table~\ref{table:1}. All panels show the magnetic field components in the radial-tangential-normal (RTN) coordinates, density, and radial bulk velocity. The RTN system is based on the position of Earth and is aligned with the line connecting the Sun and the Earth. The R axis, which is directed radially away from the Sun and passes through Earth, is the primary axis. The T axis, which is perpendicular to the R axis and parallel to the solar equatorial plane, is determined by the cross product of the Sun's spin vector and the R axis. The T axis is positive in the direction of the planet's rotation around the Sun. The N axis completes the right-handed set of axes. The blue and black lines in each graph show the OMNI data, and our simulation results at Earth for the seed ensemble member, respectively. The green lines are the simulation results for the rest of ensemble members. One can see that the uncertainties in GCS parameters can introduce quite large spreads in CME magnetic field, density, and velocity at Earth. {The vertical gray line represents the arrival time of the observed CME. The dotted red lines show the arrival time of the earliest and latest arriving ensemble members. The spread of arrival times range from as low as 12 hours in case~5 to as large as 46 hours in case~4. The arrival time of the simulated CME ensemble members is determined by comparing the simulated density at Earth between a simulation with an initialized CME and a simulation without it. The arrival time is defined as the point at which the density in the CME-initialized simulation surpasses 10\% of the density in the no CME-initialized simulation. The accuracy of this method is confirmed through visual inspection of the simulated density at Earth.}

In Cases 3, 4, and 5, the CME density at Earth is reproduced reasonably well. The results for cases 1 and 6 considerably overestimate densities at Earth. In case 2, our simulations were not able to reproduce the observed large density in the CME sheath region. The reason for disagreements in Cases 1,2, and 6 can be that we insert only 0.1\% of the total observed CME mass into the flux rope. The effect of this choice on the simulated CME densities at Earth needs to be further investigated. The plasma speeds in the simulated CMEs show a reasonable agreement with observations in all considered cases. The ensemble members show quite a large arrival time window for all six cases. For example, the arrival time window of the 12 July 2012 CME (case 5) is 12 hours wide. The rest of the 5 CMEs also had arrival time windows wider than 10 hours. In this study, we focus on the improvement of the arrival time errors. This is done by comparing the results obtained using our ensemble members with the STEREO heliospheric imager data. 

Since our CME model involves a flux rope, the modifications our simulations impose onto the magnetic field at Earth resemble magnetic clouds with smoothly rotating magnetic field components. A visual inspection shows that our model underestimates the magnetic field magnitude at Earth for cases 1, 2, 4, and 6. Magnetic field magnitude was reproduced well in cases 3 and 5. Direction of the CMEs in both these cases was most aligned with the Sun-Earth line (see latitude and longitude values in Table~\ref{table:1}. This suggests that our model underestimates magnetic field at its flanks. One of the reasons for this can be due to the reconnection of the cloud magnetic field with the SW magnetic field. This effect deserves further investigation which will be done in the future. Our ensemble modeling shows that the GCS uncertainties can introduce large changes in the CME magnetic field at Earth. CME magnetic field magnitude and direction can have a large spread in ensemble simulations. It should be noted that the range is only due to the uncertainty in GCS parameters that are used to constrain the constant turn flux rope model. If additional ensemble members were to be created to represent the uncertainty in poloidal and toroidal magnetic fluxes that are used to constrain the model, we could expect to see even a larger spread in the simulated magnetic field at Earth. This demonstrates significant difficulties in the implementation of the models, such as ours, for forecasting the CME magnetic field at Earth. However, we can try to weigh the individual ensemble members using their agreement with the heliospheric imager data to provide a probabilistic forecasting. This will be done in the future. Here we decided to concentrate on improvement of the arrival time predictions. 

\subsection{Comparison with HI data}
\begin{figure}
\centering
\center
\begin{tabular}{c c} 
\quad \includegraphics[scale=0.1,angle=90,height=6cm,keepaspectratio]{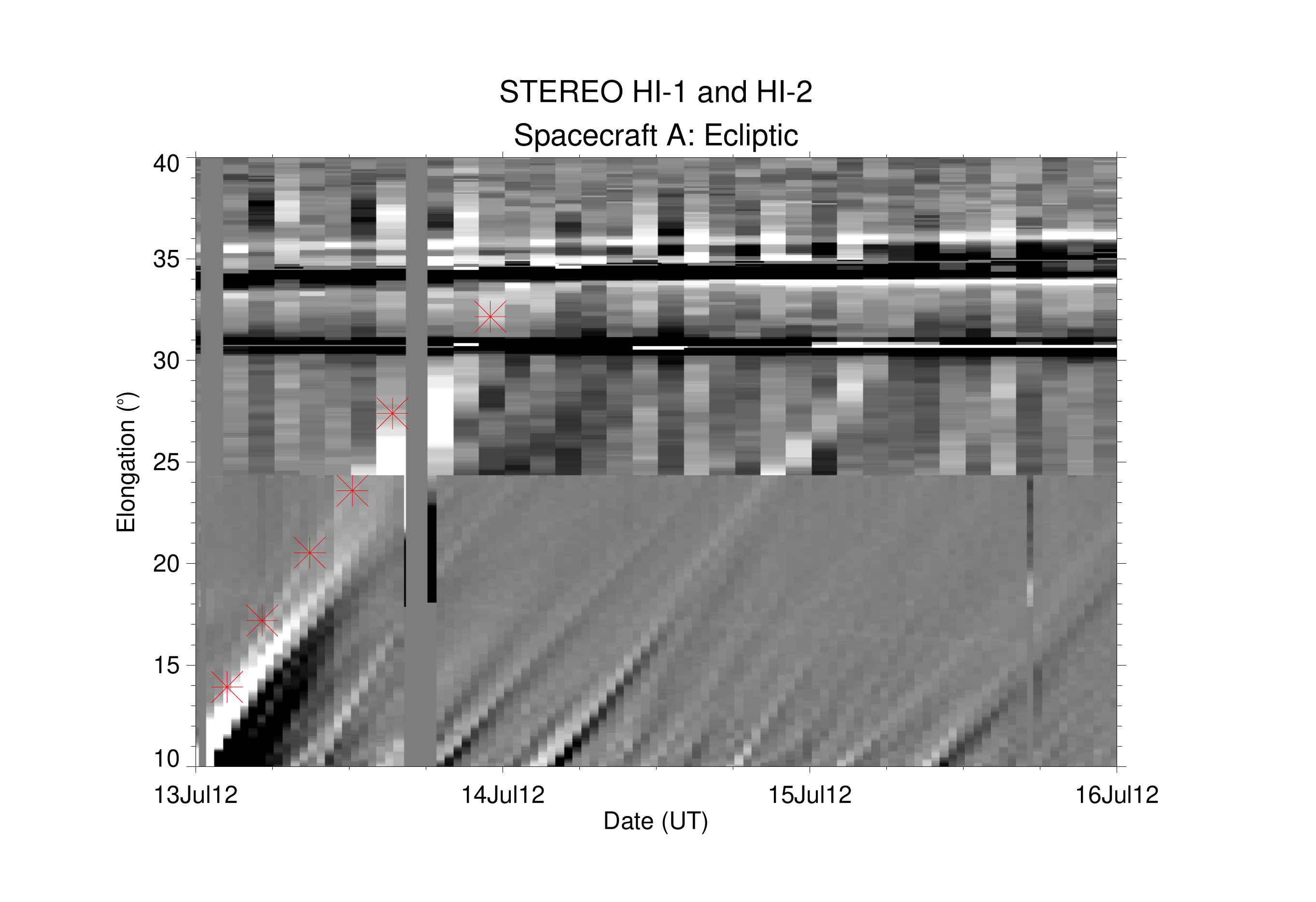} 
\includegraphics[scale=0.1,angle=90,height=6cm,keepaspectratio]{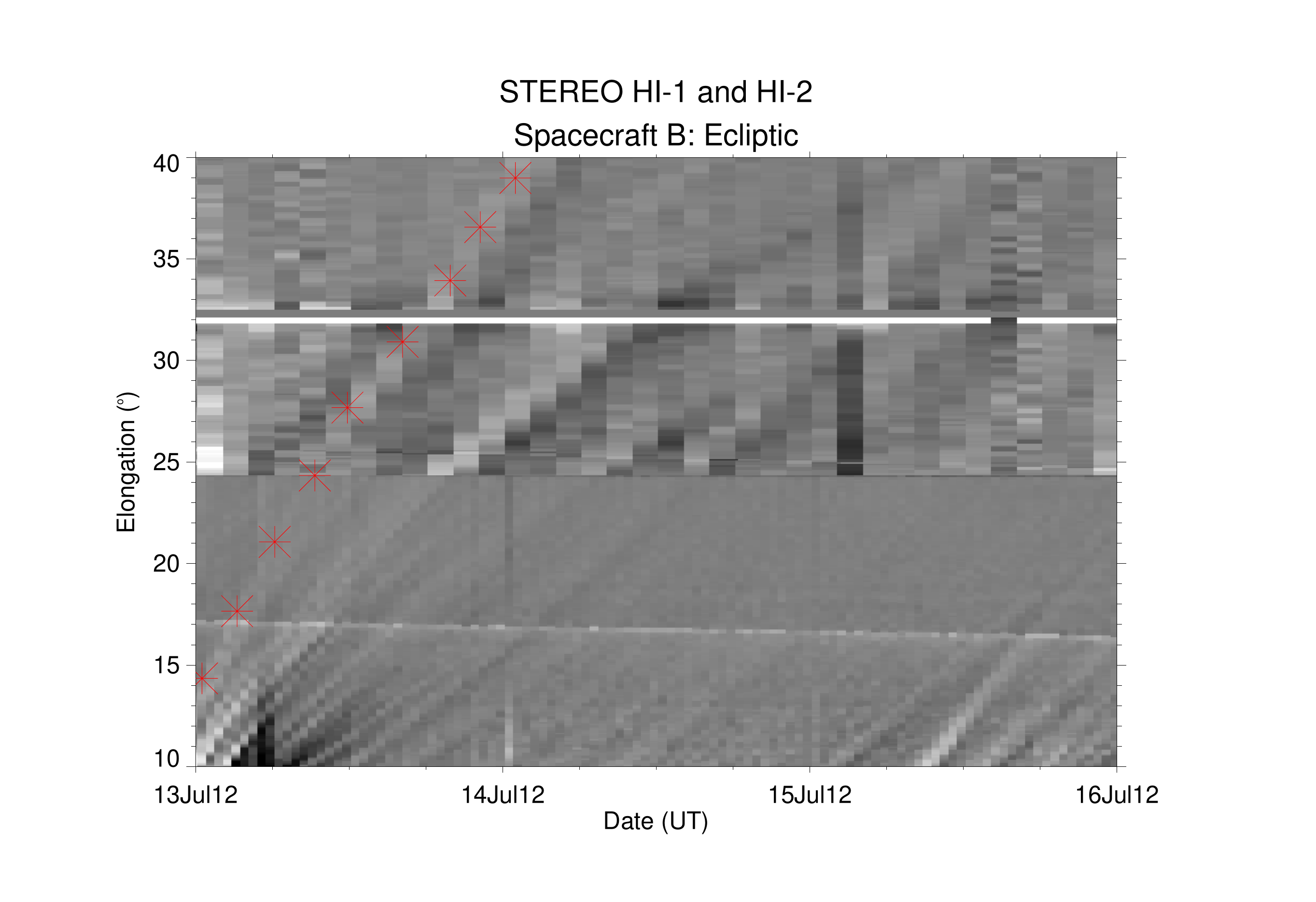} \\
\includegraphics[scale=0.1,angle=0,height=5cm,keepaspectratio]{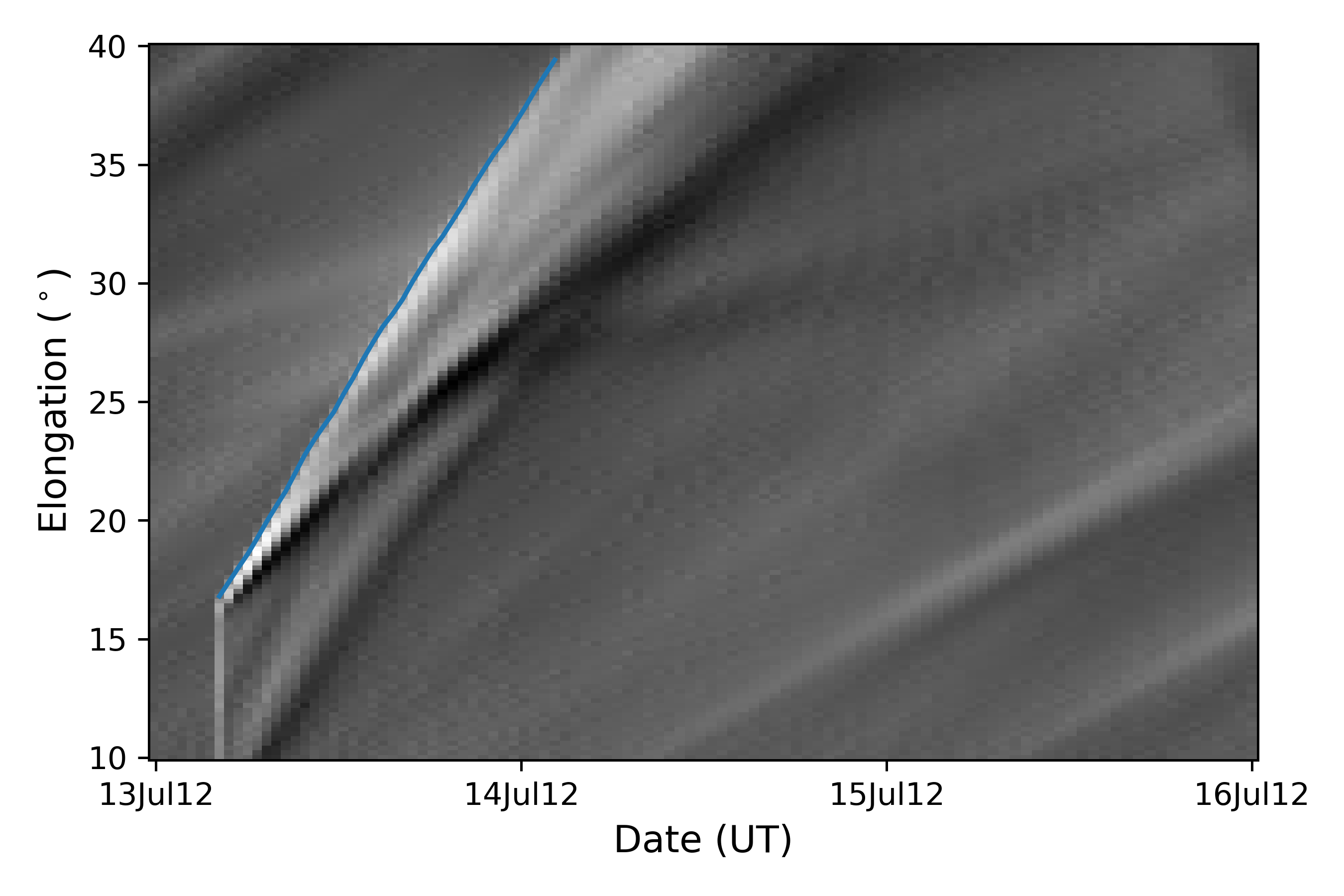} \quad \qquad
\includegraphics[scale=0.1,angle=0,height=5cm,keepaspectratio]{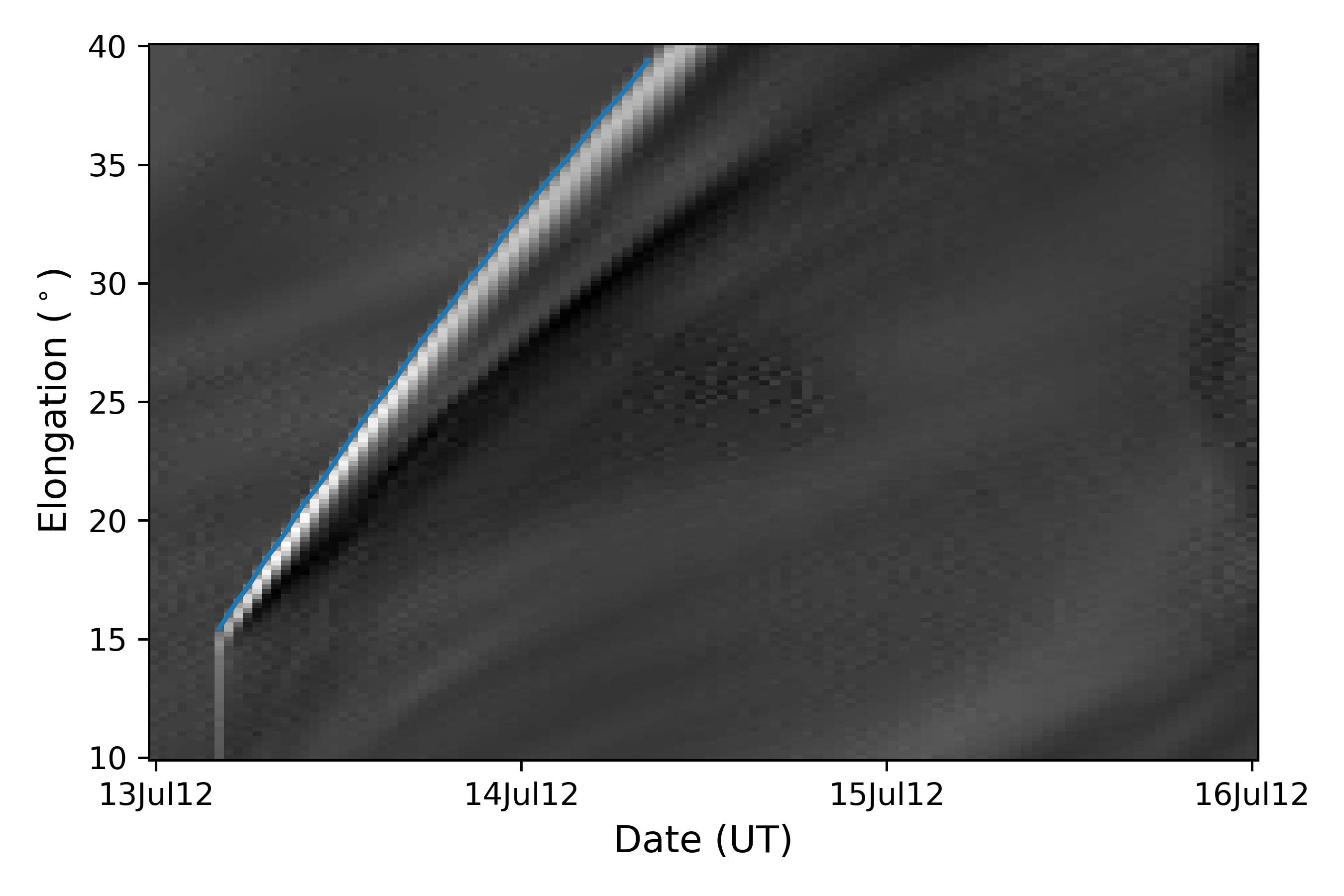}\\
\begin{overpic}[scale=0.1,angle=0,width=6cm,keepaspectratio]{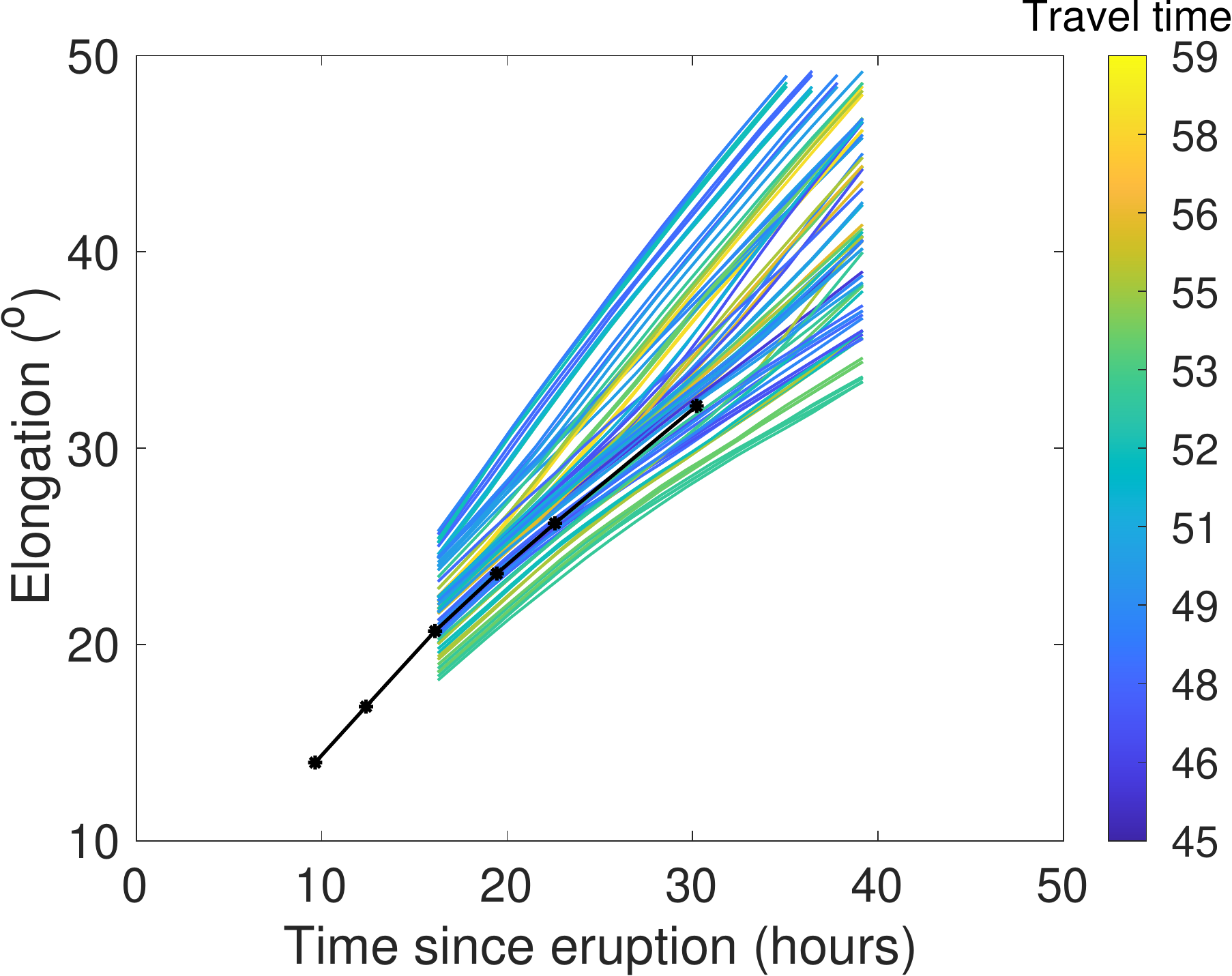}
\put(12,65){\color{black}{ \fontsize{8}{9}\selectfont 5-A}}
\end{overpic} \qquad \qquad
\begin{overpic}[scale=0.1,angle=0,width=6cm,keepaspectratio]{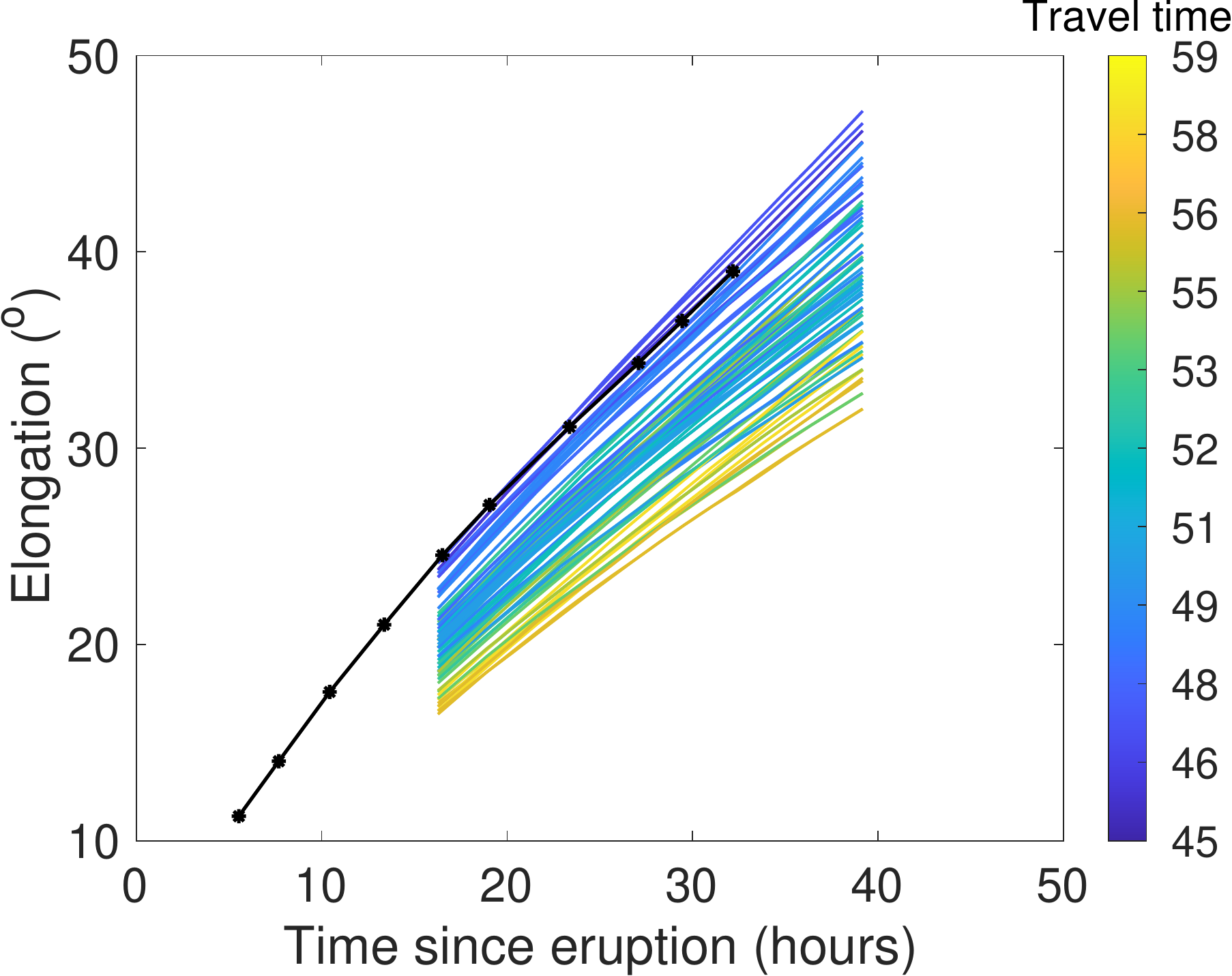}
\put(12,65){\color{black}{ \fontsize{8}{9}\selectfont 5-B}}
\end{overpic}
\end{tabular}
\caption{\textit{Top panel:} J-maps created using HI1  and HI2 data from STEREO-A (\textit{left}) and STEREO-B (\textit{right}) showing the 12 July 2012 CME (Case 5). {The tracked CME front is shown with red symbols in both J-maps}.  \textit{Middle panel:} Synthetic J-maps created from the simulation of the seed ensemble member for the 12 July 2012 CME from STEREO-A (\textit{left}) and STEREO-B (\textit{right}) vantage point. The CME front is traced with the blue line. \textit{Bottom panel:} The time-elongation data extracted from the J-maps by tracing the CME fronts in observed J-map (black line) and synthetic J-maps for all ensemble members (colored according to the travel time in hours from eruption to Earth impact). The left panel is for STEREO-A and the right panel is for STEREO-B. These panels are tagged with ``case number - spacecraft" format tags.}
\label{HI_Jmaps}
\end{figure}

{HI1} imagers on board STEREO-A \& B have fixed FOVs that extend between 4$^/circ$ and 24$^/circ$ of elongation angle. {HI2}s similarly have FOVs that extend between 18.7 and 88.7 degrees of elongation angles. {This translates to distances of up to 1 AU and beyond for {HI1} and {HI2} combined}. CMEs expand in the IH and consequently their density decreases as they propagate. This means that CMEs become fainter as they propagate in the HI FOVs, so that most of CMEs become untraceable near the outer edge of the {HI2} FOV. A common approach to tracking CMEs in HI FOV is to create time-elongation maps, also known as J-maps. These are created by extracting the {pixels corresponding to the ecliptic plane} in running-difference HI images. These extracted {pixels} are then stacked vertically for a specified time range to create an image called a J-map. CME fronts reveal themselves as oblique bright fronts after this procedure. The top panels of Figure~\ref{HI_Jmaps} show such J-maps for the 12 July 2012 CME created using STEREO-A \& B data. The 12 July 2012 CME can be seen clearly in both STEREO-A \& B J-maps. We can trace the bright fronts of a CME in both J-maps to obtain time-elongation data. J-maps can have multiple CMEs visible in them, but the bright front corresponding to the CME under consideration can easily be found by considering the eruption time of a CME. 

\begin{figure}
\centering
\center
\begin{tabular}{c c} 
%\hspace{-0.6pc}
\begin{overpic}[scale=0.1,angle=0,width=5cm,keepaspectratio]{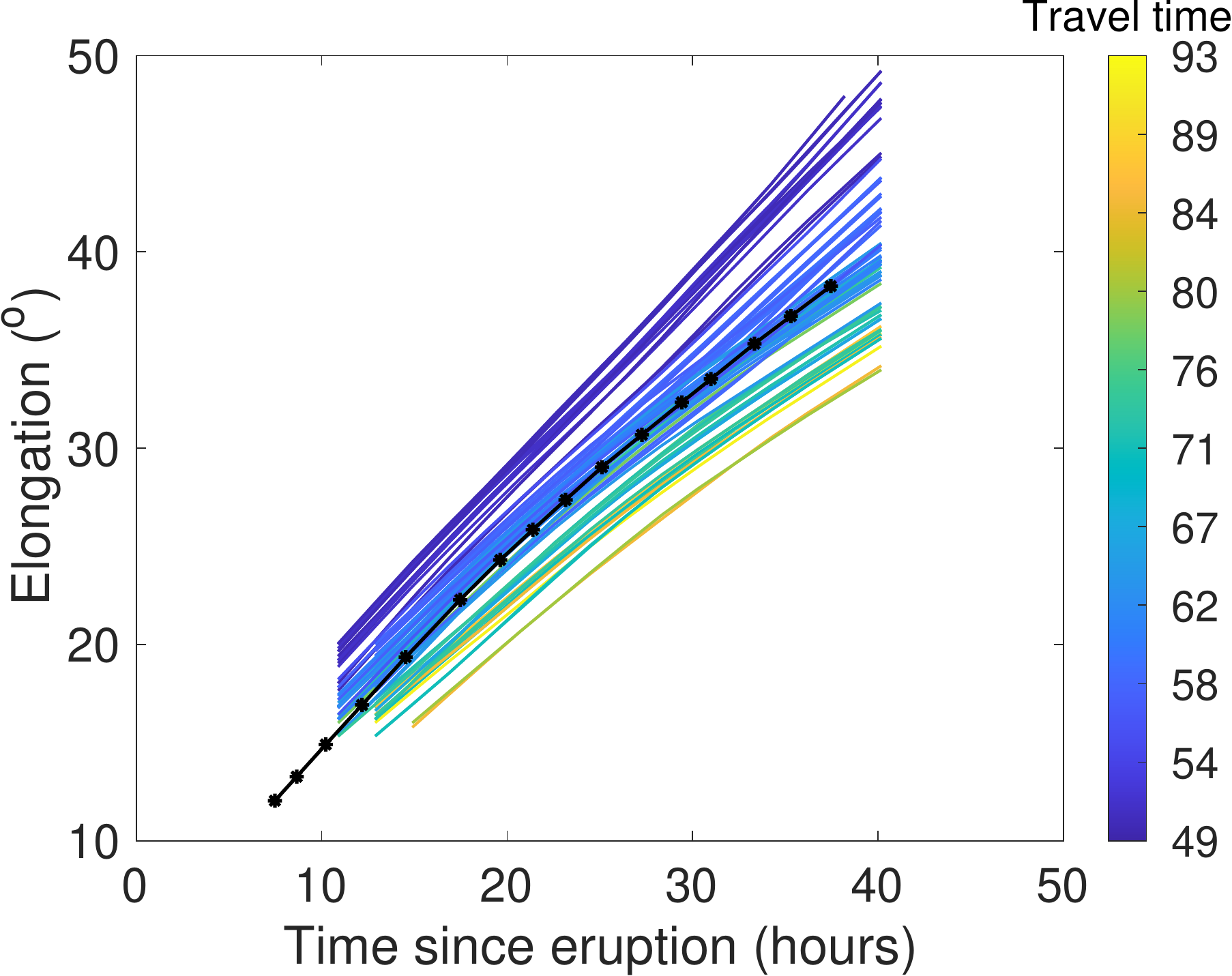}
\put(12,65){\color{black}{ \fontsize{8}{9}\selectfont 1-A}}
\end{overpic}
\begin{overpic}[scale=0.1,angle=0,width=5cm,keepaspectratio]{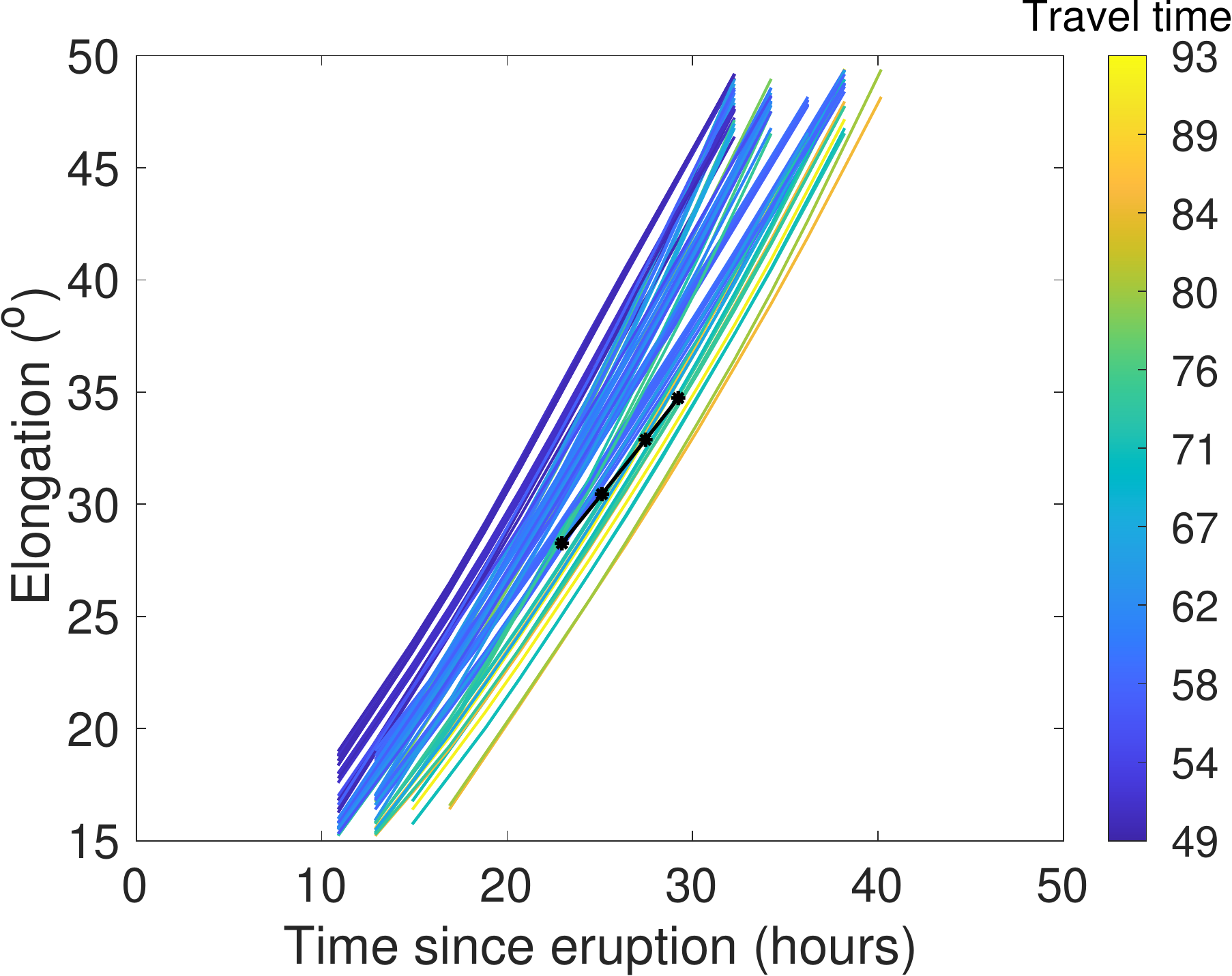}
\put(12,65){\color{black}{ \fontsize{8}{9}\selectfont 1-B}}
\end{overpic}\\
%\hspace{-0.6pc}
\begin{overpic}[scale=0.1,angle=0,width=5cm,keepaspectratio]{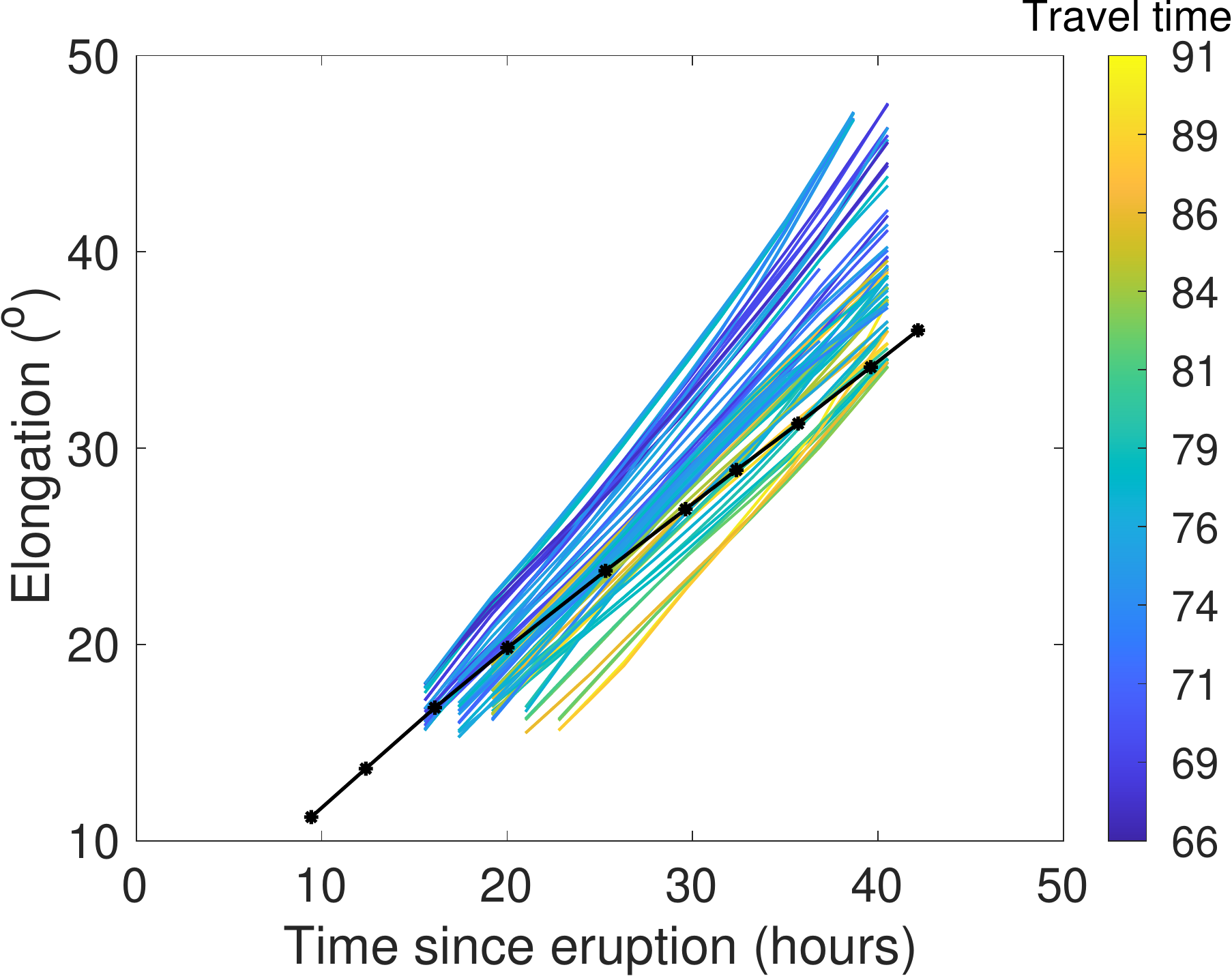}
\put(12,65){\color{black}{ \fontsize{8}{9}\selectfont 2-A}}
\end{overpic}
\begin{overpic}[scale=0.1,angle=0,width=5cm,keepaspectratio]{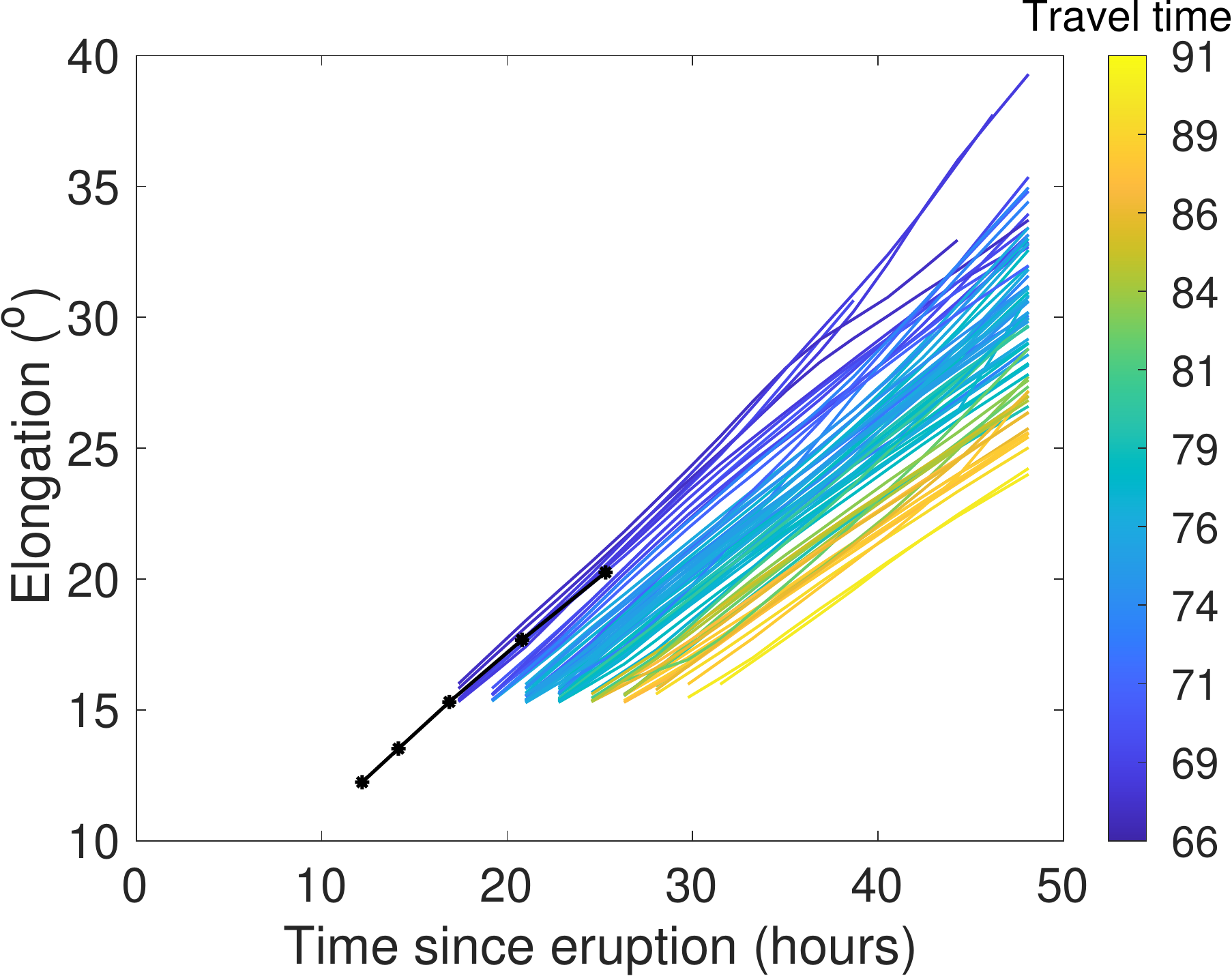}
\put(12,65){\color{black}{ \fontsize{8}{9}\selectfont 2-B}}
\end{overpic}\\
\begin{overpic}[scale=0.1,angle=0,width=5cm,keepaspectratio]{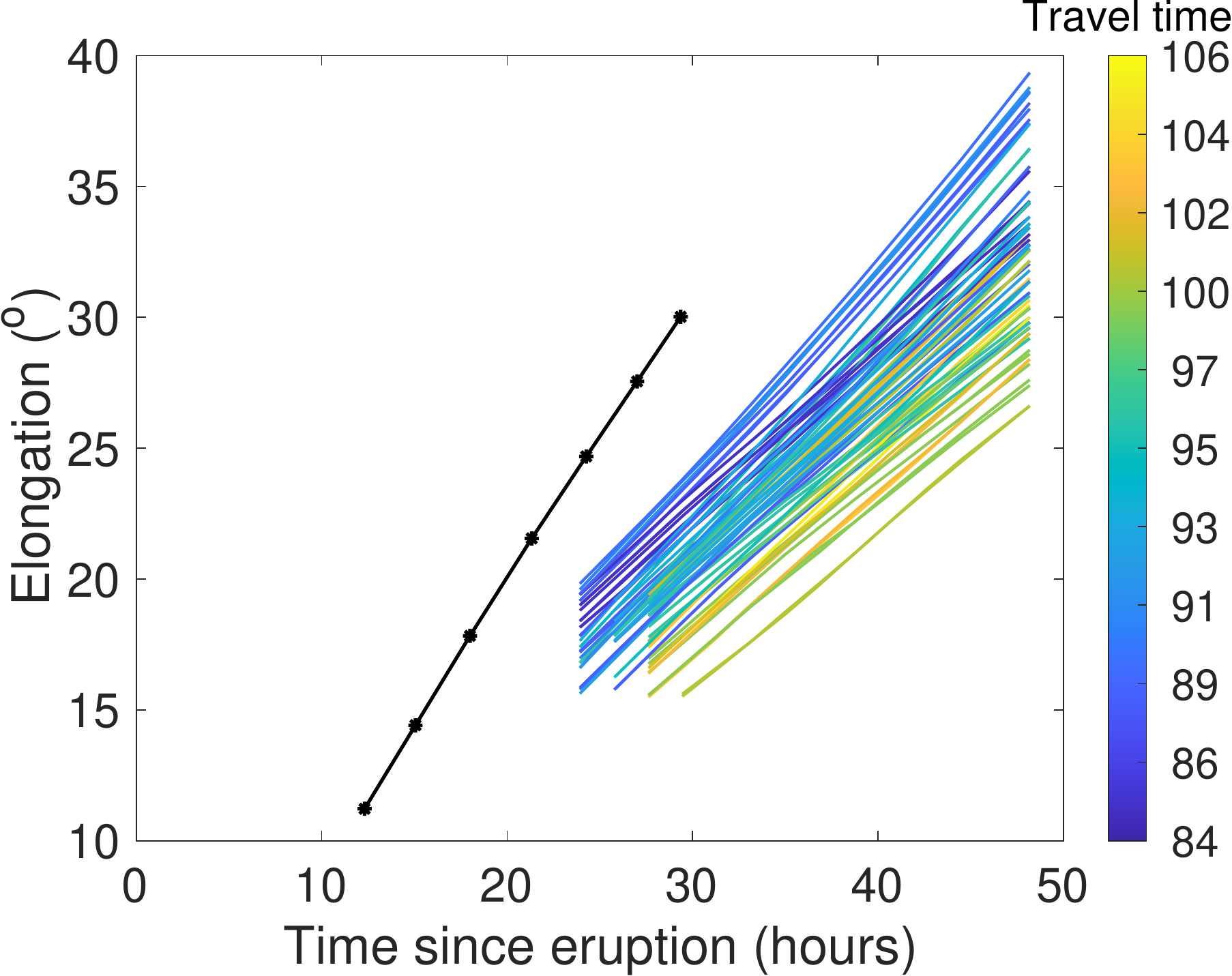}
\put(12,65){\color{black}{ \fontsize{8}{9}\selectfont 3-A}}
\end{overpic}
\begin{overpic}[scale=0.1,angle=0,width=5cm,keepaspectratio]{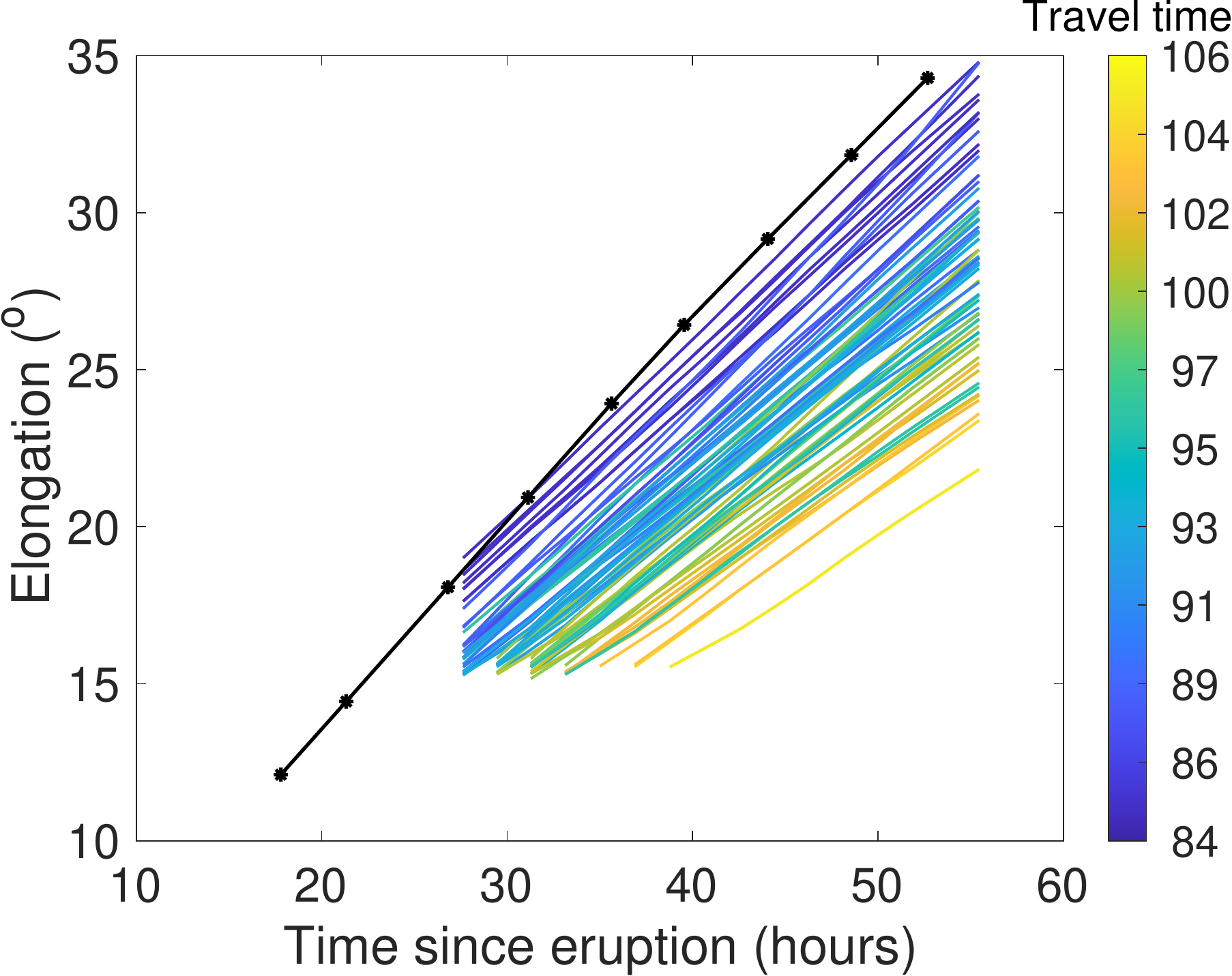}
\put(12,65){\color{black}{ \fontsize{8}{9}\selectfont 3-B}}
\end{overpic}\\
\begin{overpic}[scale=0.1,angle=0,width=5cm,keepaspectratio]{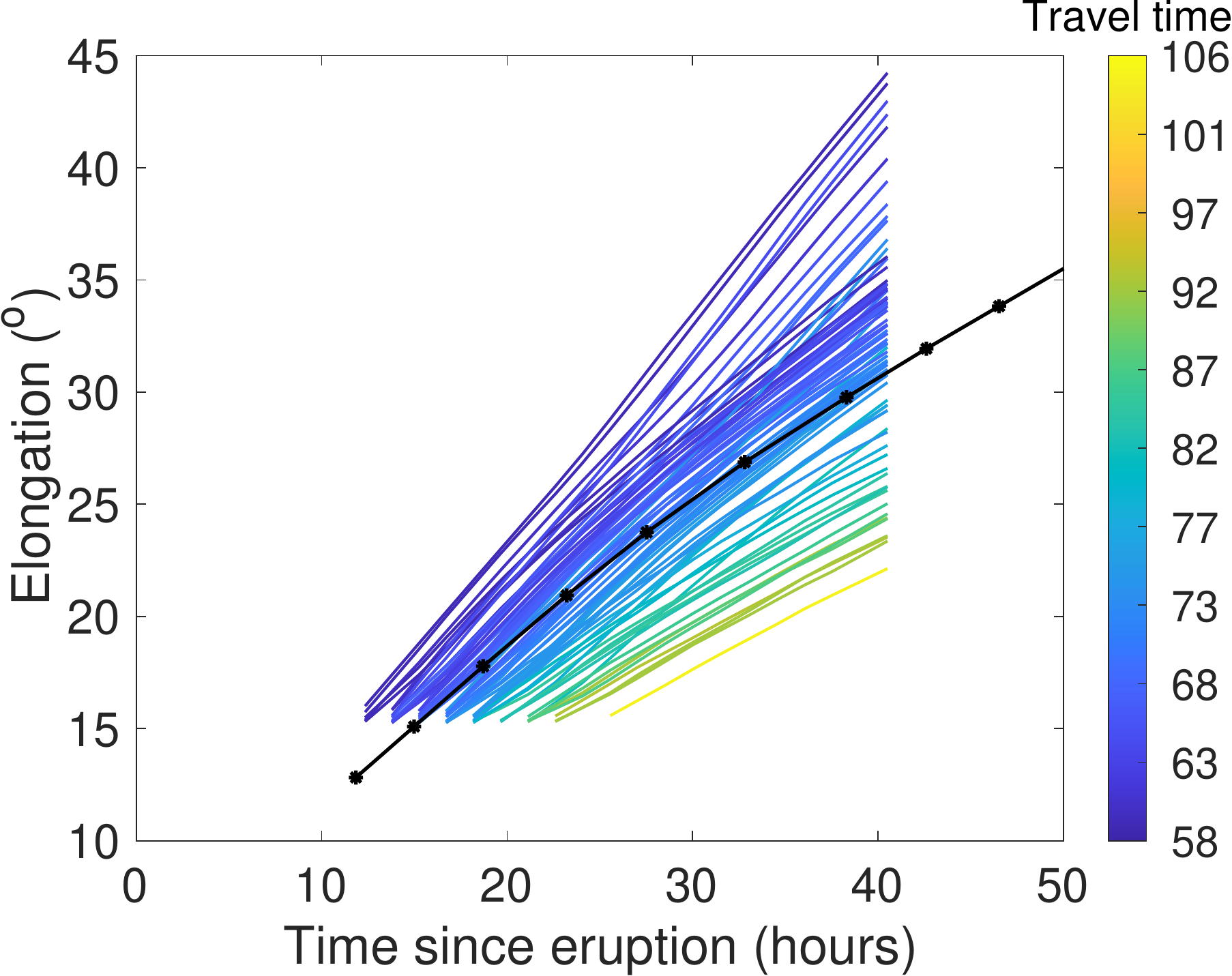}
\put(12,65){\color{black}{ \fontsize{8}{9}\selectfont 4-A}}
\end{overpic}
\begin{overpic}[scale=0.1,angle=0,width=5cm,keepaspectratio]{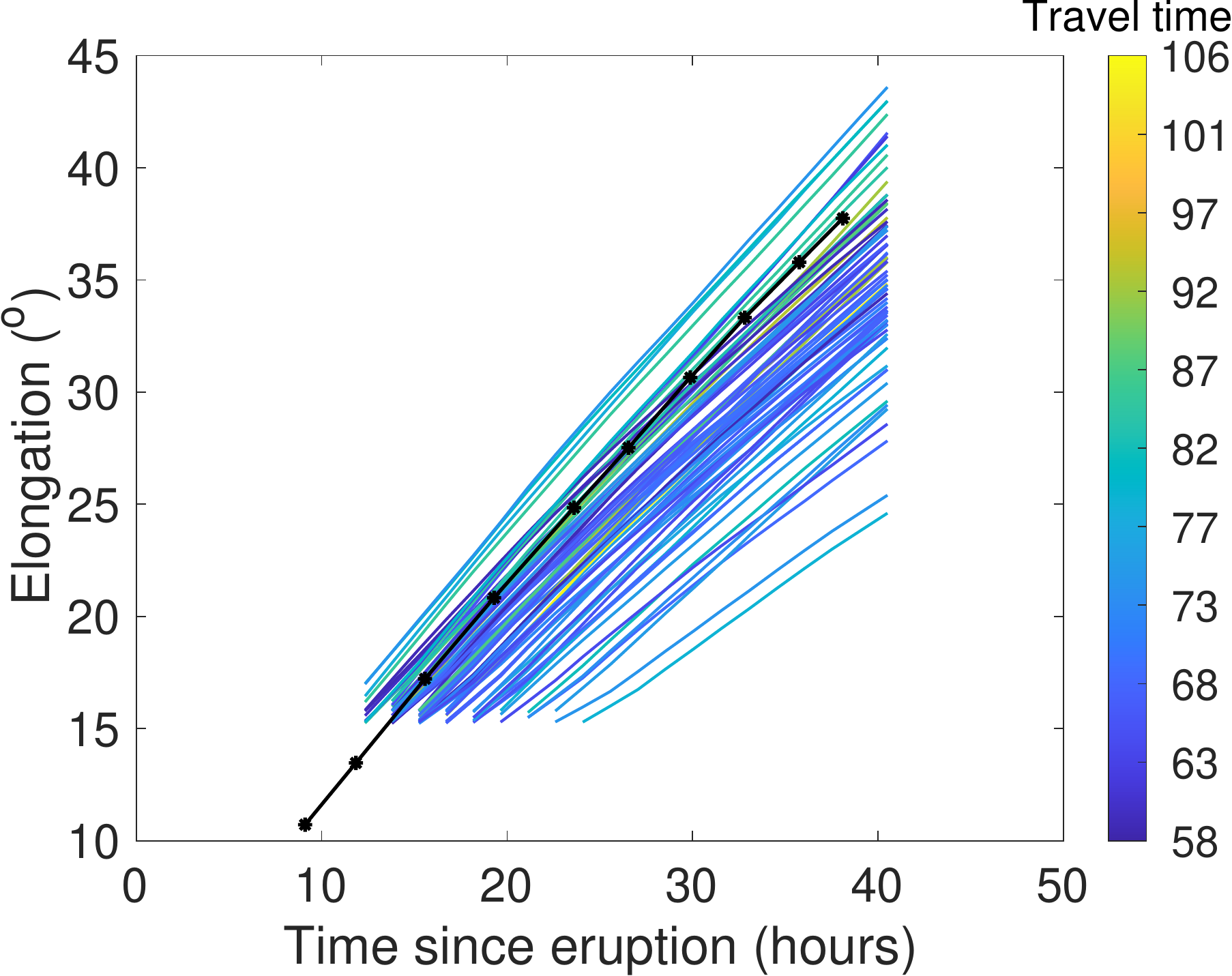}
\put(12,65){\color{black}{ \fontsize{8}{9}\selectfont 4-B}}
\end{overpic}\\
\begin{overpic}[scale=0.1,angle=0,width=5cm,keepaspectratio]{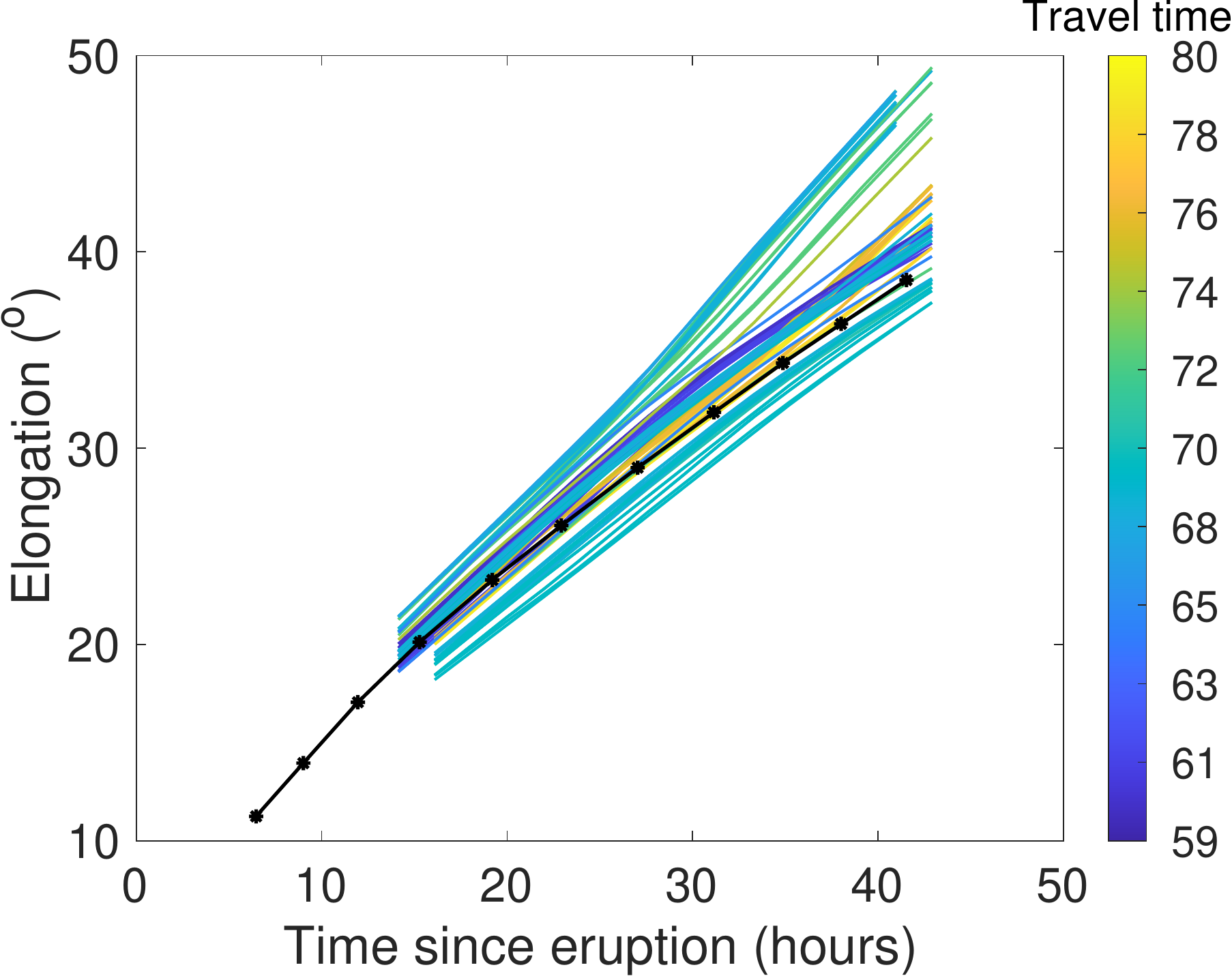}
\put(12,65){\color{black}{ \fontsize{8}{9}\selectfont 6-A}}
\end{overpic}
\begin{overpic}[scale=0.1,angle=0,width=5cm,keepaspectratio]{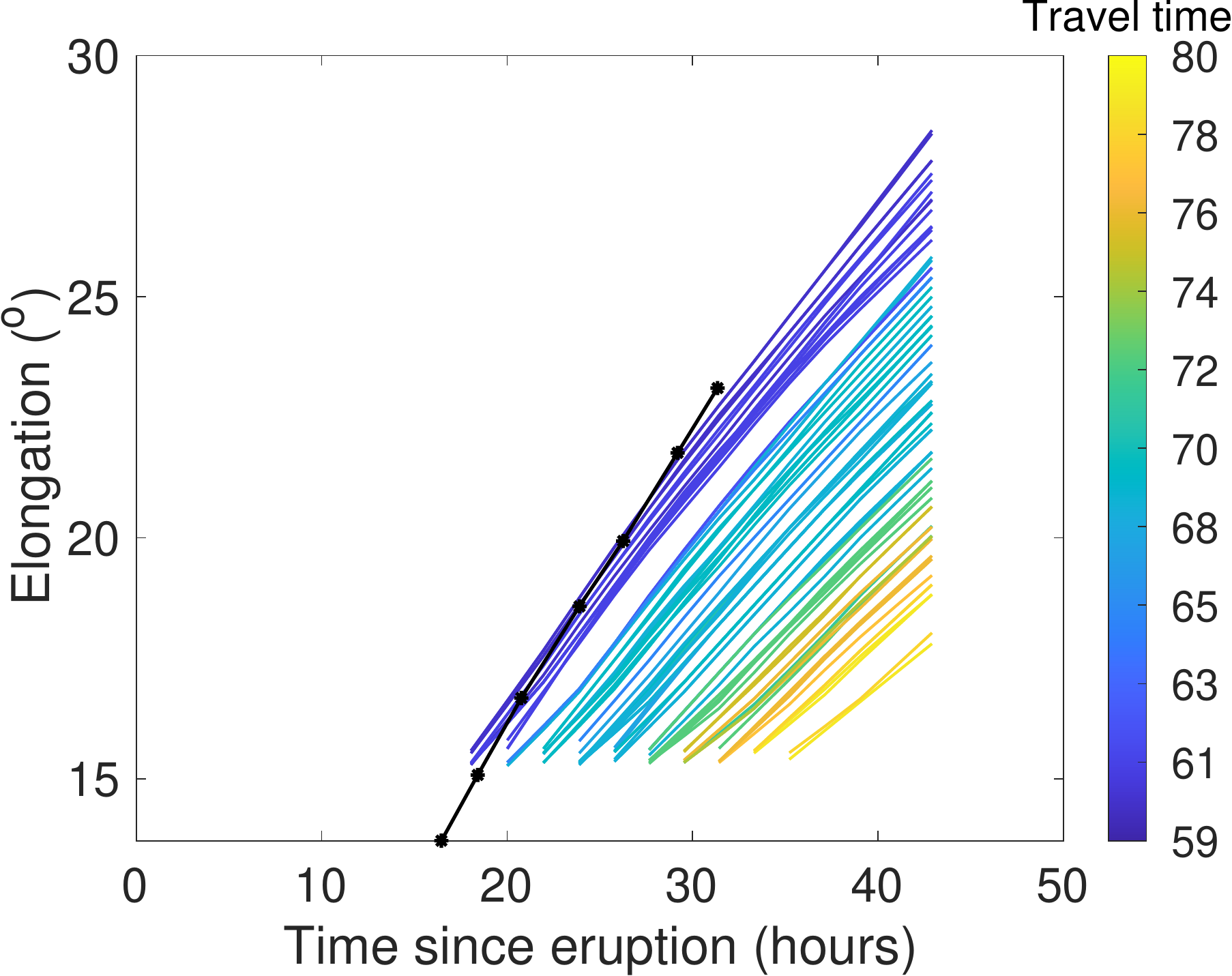}
\put(12,65){\color{black}{ \fontsize{8}{9}\selectfont 6-B}}
\end{overpic}
\end{tabular}
\caption{The time-elongation data obtained by tracing the CME fronts in the observed (black line) and synthetic J-maps for all ensemble members (colored according to the travel time in hours from eruption to the Earth impact). The left panel is for STEREO-A and the right panel is for STEREO-B. The panels are tagged with the ``case number - spacecraft" format tag in the top left corners. {Case~5 has been shown previously in Figure~\ref{HI_Jmaps}}}
\label{Jmaps_5}
\end{figure}

Simulations can be compared with observations by creating synthetic J-maps. The middle row in Figure~\ref{HI_Jmaps} shows the synthetic J-maps we have created with the simulation based on the seed ensemble member for the 12 July 2012 CME (for STEREO-A on the left and STEREO-B on the right). We create these maps by integrating the density along the line-of-sight (LOS) towards STEREO in the ecliptic plane and taking a running difference with time. We then track the bright front in the synthetic J-maps to get time-elongation data. We have automated this tracking because of the large number of ensemble members. The time-elongation data in simulations and observations can be easily compared. The bottom panel of Figure~\ref{HI_Jmaps} shows the time-elongation graphs for observations (black line) and our simulated ensemble members (color lines). The plotlines corresponding to the different ensemble members are colored according to the travel time of CME from Sun to Earth for that ensemble member, which is found by subtracting the eruption time of the 12 July 2012 CME from the arrival time for that ensemble member. Similar comparison plots for the rest of 5 cases are shown in Figure~\ref{Jmaps_5}. We can see that, for all our cases, some ensemble members agree better than others with the CME observations in the IH. The agreement can be better visualized by creating plots that show the difference between simulated and observed elongations as function of time, Figure~\ref{elon_diff} shows these plots for the 12 July 2012 CME. In the next subsection, we will describe the methods to use these comparisons to improve the arrival time estimates. 

\subsection{Improving arrival times with ML}
We have simulated the seed ensemble members for each of our six CMEs with the properties listed in Table~\ref{table:1}. The errors in arrival times for these simulations are given in Table~\ref{table:2}. Using the errors $\Delta t_i$ (in case $i$), we can calculate the mean error (ME), MAE, and standard deviation (SD) as follows:
$$\textup{ME} = \frac{1}{N} \sum_{i=1}^N \Delta t_i$$

$$\textup{MAE} = \frac{1}{N} \sum_{i=1}^N |\Delta t_i|$$

$$\textup{SD} = \sqrt{\frac{1}{N}\sum_{i=1}^N |\Delta t_i - \textup{ME}|^2}$$

The MAE obtained by simulating only the seed ensemble members, for all six cases, is $\sim$8 hours. All simulated CMEs in the seed members arrived later than in the observations. Therefore the ME has the same magnitude as MAE, but the sign is negative. This shows that our CME simulations tend to have a negative bias in the arrival time estimates. Cases 2, 3, and 6 have arrival time errors equal to $-15$ hours, $-15.2$ hours, and $-9.1$ hours and their contribution to the MAE is therefore dominant. These three cases also have the launch speeds lower than the other three CMEs. This shows the tendency of our model to have large arrival time errors for slower CMEs.

\begin{figure}
\centering
\center
\begin{tabular}{c c} 
\begin{overpic}[scale=0.1,angle=0,width=6cm,keepaspectratio]{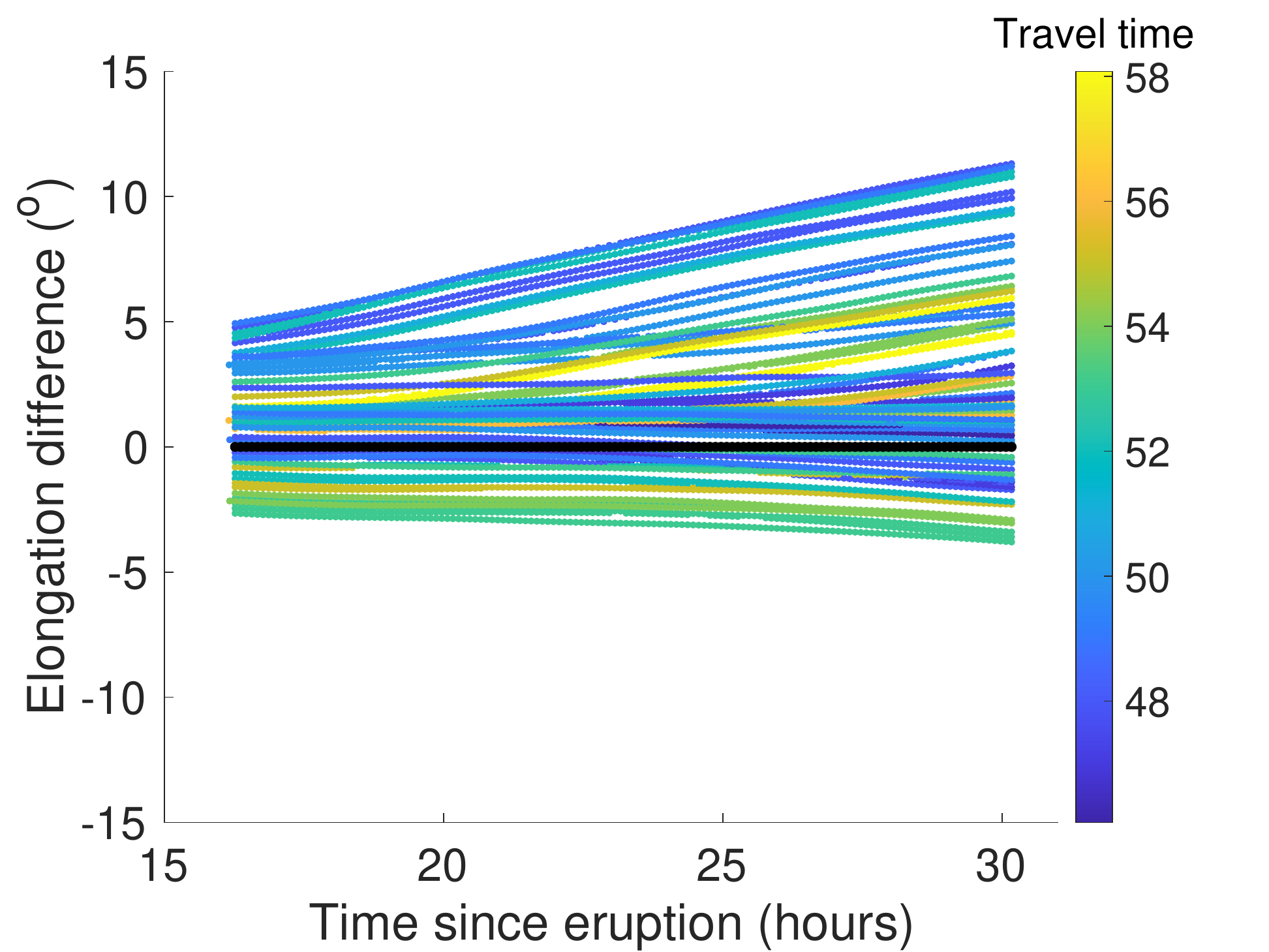}
\put(15,65){\color{black}{ \fontsize{8}{9}\selectfont 5-A}}
\end{overpic}
\begin{overpic}[scale=0.1,angle=0,width=6cm,keepaspectratio]{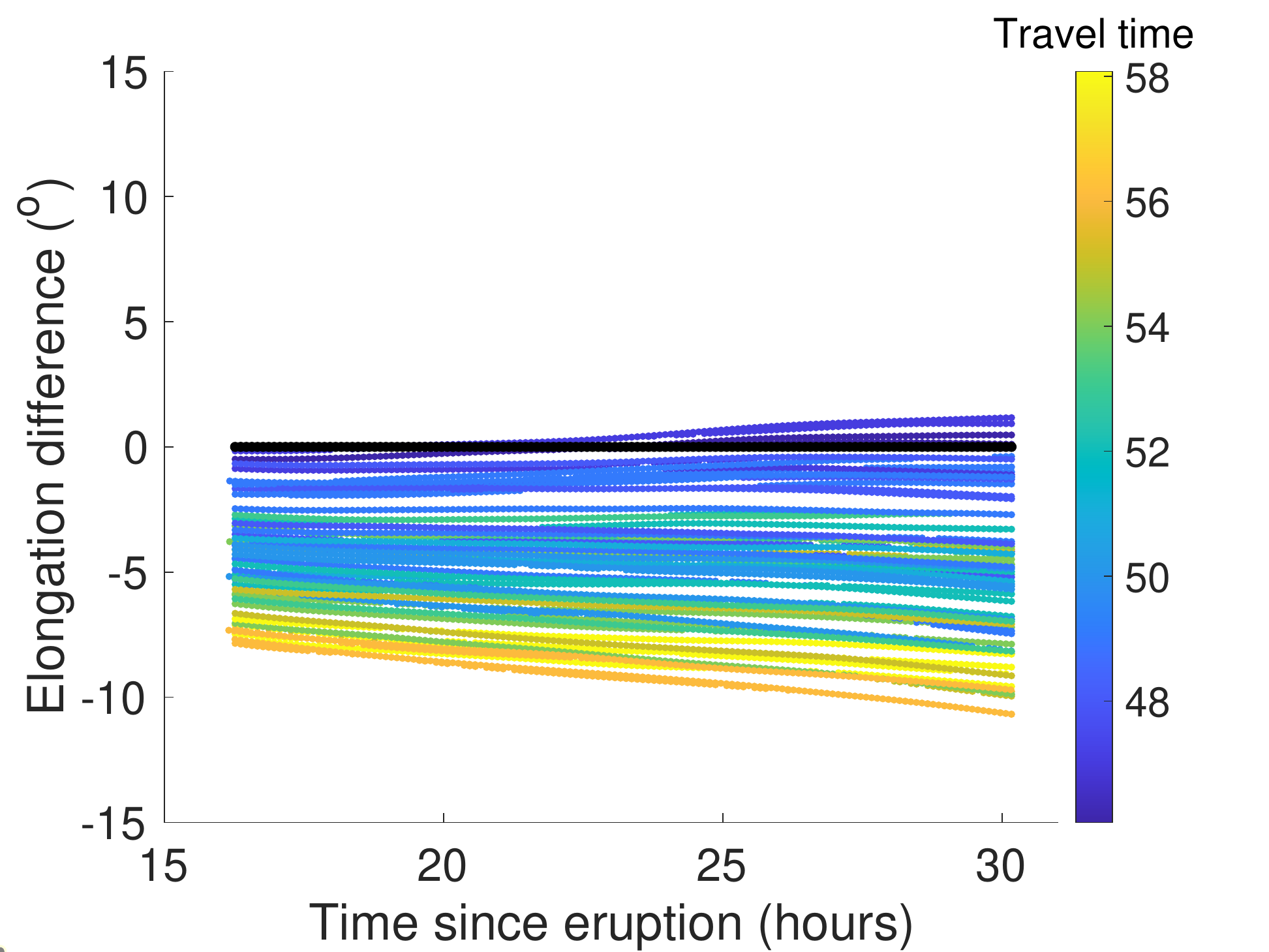}
\put(15,65){\color{black}{ \fontsize{8}{9}\selectfont 5-B}}
\end{overpic}
\end{tabular}
\caption{The difference of the elongation angles obtained from the synthetic and observed J-maps for 12 July 2012 CME for STEREO-A (\textit{left panel}) and STEREO-B (\textit{right panel}). The panels are tagged with the ``case number - spacecraft" format tag in the top left corners.}
\label{elon_diff}
\end{figure}

MAE is a more meaningful metric than ME to gauge the performance of a model, since ME calculations are affected by the mutual cancellation of positive and negative $\Delta t_i$ \citep{Riley2018}.  We can further reduce MAE by using the arrival time distribution for all ensemble members and their corresponding differences between the simulated and observed elongation angles in the IH. We can achieve this by answering the following question: what will be the arrival time in a CME simulation of hypothetical ensemble member whose elongation angle evolution in its synthetic J-map matches perfectly with the observations in the IH? This hypothetical ensemble member can be represented by the ``elongation difference = 0" line in plots shown with the black line in Figure~\ref{elon_diff}.

We have used two machine learning-based methods to answer this question: (1) the lasso regression (LR) \citep{Tibshirani} and (2) the fully connected neural networks (NNs). In a machine learning approach, a hypothesis is formulated to map the input variables onto an output variable with the goal that the algorithm learns the mapping of a function $h:X-y$ such that $h(x)$ is a good approximation for the target variable $y$. Training of the algorithm involves the minimization of the cost/loss function for our hypothesis.  To reduce over-fitting, penalties are added to the cost function to avoid large coefficients which make the model unstable. The method of adding penalties to the cost function is known as regularization. 

{LR, also known as L1 Regularization, is a linear regression variant that adds a penalty to the loss function to reduce the magnitude of the coefficients for the features or inputs to the model. The penalty term is the absolute value of the coefficients multiplied by a scalar constant, called the regularization parameter. The objective function to be minimized in LR is the mean squared error plus the penalty term, which is defined as follows:}
$$J(w)=\frac{1}{N}\sum_{i=1}^{N}(y_i-\sum_{j=1}^{p}w_jx_{ij})^2 + \lambda \sum_{j=1}^{p}|w_j|,$$
{where $w$ is the vector of coefficients, $N$ is the number of samples, $p$ is the number of inputs/features, $x_{ij}$ is the $j$th feature of the $i$th sample, $y_i$ is the target variable, and $\lambda$ is the regularization parameter that controls the amount of shrinkage of the coefficients. The LR forces some of the coefficients to be exactly zero, resulting in the elimination of some features. This means that only the most important features are included in the model, reducing the risk of overfitting.}

{Fully connected networks are based on the idea of artificial neurons, which are modeled after biological neurons in the human brain. These artificial neurons receive inputs from other neurons, process these inputs through a non-linear activation function, and produce an output that can be used as input for other neurons in the network. By stacking multiple layers of neurons, a fully connected network can learn complex representations of data, allowing it to make accurate predictions and decisions.} Neural networks can approximate a function by calculating the weighted sum of inputs, including the bias term, computing the activation function response, and passing the output to the output neuron. This constitutes the forward pass or the forward propagation. The cost/loss function is calculated by ``back propagation'' of the cost function gradients \citep{Rumelhart}. Machine learning techniques, including the NNs, have been successfully applied to prediction and forecasting problems in heliophysics \citep{Borda, QuML, Li_Wang, Qahwaji, Wang_Cui, Yuan_Shih, Ahmed_Qahwaji, Bobra_Couvidat, Jonas, Benson_1, Benson_2, Benson_3}. \cite{Enrico} provides an extensive review of the methods and applications of machine learning research in heliophysics.  

\begin{figure}[h]
\center
\includegraphics[scale=0.8,angle=0,keepaspectratio]{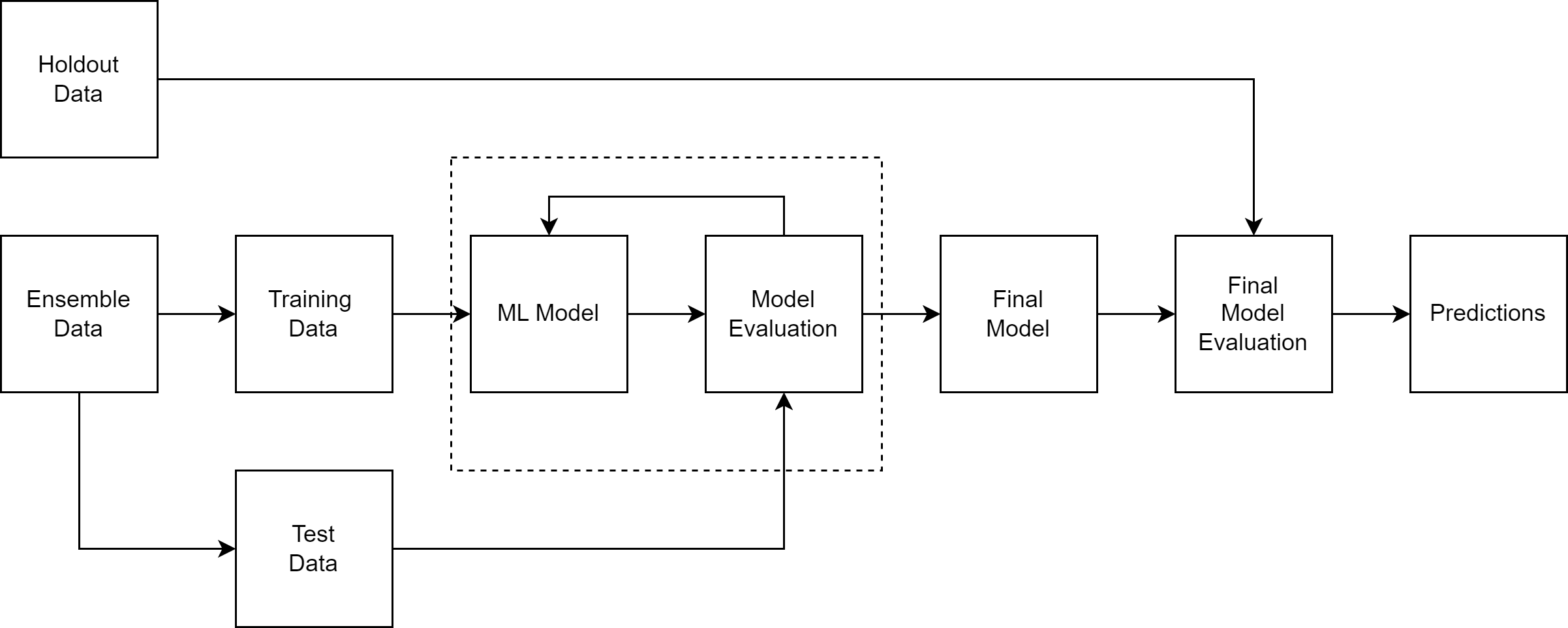}
\caption{{Training and evaluation pipeline for machine learning algorithms. ``Ensemble Data" contains independent variables: time since the eruption, and elongation angle difference between observed and simulated CMEs, and the dependent variable: travel time of the simulated CME to reach Earth. In ``Holdout Data", elongation angle difference is set as zero at all times since CME eruption. ``Predictions" are the travel time and it's standard deviation for the given ``Holdout Data".}}
\label{ml_process}
\end{figure}

For both methods, the variables, such as time since the eruption, elongation angle difference from STEREO-A, and elongation angle difference from STEREO-B ({as shown in Figure~\ref{elon_diff}}), are considered to be independent, whereas the travel time for an ensemble member is considered to be the dependent variable. {These independent and dependent variables constitute the ``Ensemble Data'' in Figure~\ref{ml_process}, which shows the model training and evaluation pipeline for the machine learning experiments. The input ensemble data is divided into training and test sets using a 80\%-20\% split. The ML model (LR or NN) is trained using the training data and the errors in predictions are evaluated using the test set. This process is iterated until we find the best model fit. The final model is then evaluated using the holdout data to make the final predictions of the CME travel time.} 

For the LR method, we frame the arrival time prediction as a simple regression problem in which we are predicting a continuous variable for a given set of inputs. The standard algorithm for regression problems is linear regression \citep{Montgomery} where the input variables are assumed to depend linearly on the target variable. The experiments using LR are implemented using the widely available Scikit-Learn library \citep{scikit-learn}. We train our model for each CME separately. The model is provided the arrays of time since eruption and elongation angle differences as dependent variables. The travel time of an ensemble member is considered to be a dependent variable. The model is tested on a holdout data set which has ``elongation difference = 0" for all times since eruption. The model is therefore determining the travel time which it expects from a hypothetical ensemble member which matches perfectly with observations in IH.

Similarly, we use a fully connected NN with twelve thousand parameters using the same input parameters to predict the CME arrival time.
The NN consists of an input layer, four hidden layers with 128, 64, 32, and 16 fully connected neurons, and a single output layer that outputs a continuous variable. A ReLU activation layer is used after every layer with batch normalization and a dropout of 20\% is used as means of regularization to avoid over-fitting. All experiments involving NNs are implemented using Tensorflow 2.0 \citep{Tensorflow}. The data for each CME are split into training, testing, and validation sets using a 60\%, 20\%, and 20\% random split. The best model is then evaluated using a separate holdout set which has ``elongation difference = 0" for all times since eruption.

We have performed two experiments each using both of the ML methods. The first experiment uses data from both STEREO-A \& B both and the second experiment uses data from the STEREO-A only. The second experiment is motivated by the current operational availability of STEREO-A data only. Our aim is to understand the applicability of our approach in this case.

The results reported in Table \ref{table:2} show the MAE, ME, and standard deviation for each experiment. The results show that the LR model based on STEREO-A \& B data performs best with a CME arrival time MAE (SD) of 4.1 (3.2) hours, followed by the NN model using STEREO-A \& B data of 5.3 (6.1) hours. However, while using the STEREO-A data alone, the NN model offers better performance than the LR model with an arrival time of 5.1 (5.2) hours compared to 6.9 (5.3) hours. We can see that all our experiments have improved the MAE compared to using just seed member simulations for the six CMEs. Perhaps the most encouraging result is the MAE of 5.1 hours achieved by using the NNs and only the STEREO-A data. It shows that NNs are more capable of handling non-linear relationships in data. This result is most relevant to the current data availability of STEREO data since only STEREO-A is operational. All our experiments resulted in a reduction in arrival time errors for cases 2, 3, and 6, which were contributing most to MAE. Our work highlights the importance of combining machine learning and physics-based MHD modeling to improve space weather predictions. The machine learning models allow us to continuously update the arrival time estimates of CMEs by using simulation-to-observation comparisons in the IH. This proof-of-concept study shows that our methods can improve the arrival time estimates with sufficient lead times.

The MAE we have achieved with our methods is better than those achieved by such models as HM, SSE, and ElEvoHI, even though those models assume CME shapes that should perfectly match the HI observations. The advantage of our model is in a more realistic CME-SW interaction. This allows for CMEs to acquire complex shapes by interacting with local SW structures, e.g., high-speed streams. In contrast, the above-mentioned models assume much simpler CME geometries throughout the IH. {A newer version of ElEvoHI, called ElEvoHI 2.0 \citep{Hinterreiter} allows for a deformable front and takes CME-SW interaction into account through an empirical drag force. Encouragingly, \citet{Hinterreiter} showed an improvement in the CME arrival time estimates when a realistic CME-SW interaction and a deformable front are taken into account. This study supports our argument that MHD modeling that reproduces realistic CME-SW interactions can provide better results than the models using CME shape assumptions throughout the IH}.

{In this study, we observed that the MAE achieved was significantly lower than the average MAE of currently operational models. However, it is important to note that the cases simulated in this study were well observed, while the MAE of the operational models typically encompasses a wider range of CMEs with varying degrees of complexity, including those with CME-CME interactions. To validate our findings, we plan to expand the scope of our study in the future to include a more diverse set of CMEs. Despite the limited sample size of six CMEs in this study, the benefit of combining machine learning and MHD CME modeling to enhance arrival time predictions is quite evident.}

\begin{table}
\centering
\caption{Errors in arrival times of the six CMEs discussed in this study, alongside with the mean error (ME), mean absolute error (MAE), and standard deviation (SD) for each method.\label{table:2}}
\begin{adjustbox}{max width=1.1\textwidth,center}
\begin{tabular}{|P{2.4cm}|P{1.55cm}|P{1.75cm}|P{1.55cm}|P{1.55cm}|P{1.55cm}|P{1.55cm}|P{0.9cm}|P{0.9cm}|P{0.9cm}|} 
\hline
%CME \#  & 1 & 2 & 3 & 4 & 5 & 6 & ME & MAE & SD\\
CME & 01/08/10 & 06/09/11 & 13/09/11 & 19/01/12 & 12/07/12 & 28/09/12 & ME & MAE & SD\\
\hline
Seed member & -3.9 & -15.0 & -15.2 & -3.7 & -0.9 & -9.1 & -8.0 & 8.0 & 6.1\\ \hline
LR (A \& B) & -3.3$\pm$0.8  & -9.3$\pm$0.2 & 1.8$\pm$0.3 & -6.8$\pm$1.7 & 1.9$\pm$0.1 & -1.8$\pm$0.4 & -2.9 & 4.1 & 3.2\\ \hline
LR (A) & -7.2$\pm$0.1 & -14.8$\pm$1.3 & 0.8$\pm$0.3 & -7.9$\pm$2.1 & -2.6$\pm$0.1 & -8.0$\pm$1.3 & -6.6 & 6.9 & 5.3\\ \hline
NNs (A \& B) & -2.2$\pm$2.6 & -12.7$\pm$1.0 & 5.9$\pm$1.3 & -4.8$\pm$1.7 & -1.0$\pm$0.3& -5.2$\pm$1.3 & -3.3 & 5.3 & 6.1\\ \hline
NNs (A) & -3.9$\pm$0.2 & -10.2$\pm$1.9 & 5.4$\pm$2.1 & -5.3$\pm$1.6 & -1.1$\pm$2.2 & -4.6$\pm$5.5 & -3.3 & 5.1 & 5.2\\ 
\hline
\end{tabular}
\end{adjustbox}
\end{table}

\section{Conclusion}\label{sec:Conclusions}
In this study, we have used a constant turn flux rope model to simulate six CMEs in a data-driven SW background. The flux rope model was constrained with the observed CME direction, tilt, speed, half-angle, aspect ratio, mass, magnetic fluxes, and magnetic helicity sign. The average uncertainties in observed direction, tilt, speed, half-angle, and aspect ratio allowed us to create 76 additional ensemble members for each CME from their seed ensemble members. We have presented the simulation results for all the ensemble members for our six considered CMEs at Earth. Using synthetic and observed J-maps, we were able to compare the ensemble members with observed CMEs. These comparisons were used in the LR and NN models to improve the arrival time estimates of the CMEs. The results of this work can be summarized as follows:

\begin{enumerate}[noitemsep,nolistsep]
    \item Constant turn flux rope model is suitable to simulate CMEs in the IH.
    \item This model can simulate CMEs with a large range of speeds, magnetic fluxes, and widths. In this study, we have simulated CMEs that were launched with speeds ranging between 480~km/s and 1357~km/s, poloidal fluxes ranging between $0.9\times10^{21}$ and $14.0\times10^{21}$ Mx, and half angles ranging between $23^\circ$ and $68^\circ$.  
    \item By simulating only the seed ensemble member of each CME, we found the MAE to be 8 hours and the standard deviation is 6.1 hours. All these simulations resulted in the simulated CMEs arriving later than the observed ones, showing a negative bias in our method. 
    \item The uncertainty in GCS parameters can be used to create ensemble members and the CME arrival time estimates can be improved by comparing simulated ensemble members with STEREO HI observations.
    \item We obtained the MAE for our six CMEs as small as 4.1 hours using the LR and STEREO-A \& B data, 6.9 hours using the LR and STEREO-A data only, 5.3 hours using NNs and STEREO-A \& B data, and 5.1 hours using NNs and STEREO-A data.
    \item The standard deviation for the six CMEs was found to be 3.2 hours using the LR and STEREO-A \& B data, 5.3 hours using LR and STEREO-A data, 6.1 hours using NNs and STEREO-A \& B data, and 5.2 hours using NNs and STEREO-A data.
    \item These MAE and standard deviation measures are better than the other currently operational and proposed arrival time forecasting models. This demonstrates that the combination of HI data and MHD modeling together with machine learning can substantially improve space weather predictions.
\end{enumerate}

The simulation of approximately 6 days of physical time on a grid of $150\times128\times256$ by MS-FLUKSS requires 2 hours of computation using 128 CPUs. Given the ability to run all ensemble members simultaneously, the ensemble modeling carried out here is practical for predicting space weather in real-time. The machine learning methods we have used are quite robust and do not take more than a few minutes to run. We have used science data from HIs onboard STEREOs for this analysis. However, these data become available several days after observations. Future work will test the applicability of the beacon HI data to our methods, which can make our methods suitable for real-time forecasting. This work also highlights the importance of having HI observatories on Lagrangian points 4 and 5 to continuously image the CME evolution in IH. {The European Space Agency's Vigil spacecraft is slated to be stationed at the L5 point in the mid-2020s, equipped with HI instruments to observe CMEs in the interplanetary space. Our current research can be leveraged to make the most of these data, thereby enabling enhanced space weather forecasts.}
 
\subsection*{Acknowledgments}
The authors acknowledge support from NSF/NASA  SWQU grant 2028154. TKK acknowledges support from AFOSR grant FA9550-19-1-0027 and NASA grant 80NSSC20K1453. NP was also supported, in part, by NSF-BSF grant 2010450 and NASA grants 80NSSC19K0075 and 80NSSC21K0004. Supercomputer allocations were provided on SGI Pleiades by NASA High-End Computing Program award SMD-21-44038581 and also on TACC Stampede2 and SDSC Expanse by NSF XSEDE project MCA07S033. We acknowledge the NASA/GSFC Space Physics Data Facility's OMNIWeb for the SW and IMF data used in this study. We also used SOHO and STEREO coronagraph data from\url{ https://lasco-www.nrl.navy.mil} and \url{stereo-ssc.nascom.nasa.gov} respectively. SDO EUV and magnetogram data have been taken from \url{http://jsoc.stanford.edu/ajax/exportdata.html}. 
This work utilizes data produced collaboratively between the Air Force Research Laboratory (AFRL) and the National Solar Observatory.

\clearpage
\bibliographystyle{plainnat}
%\bibliography{Singh2021}

\end{document}